\documentclass[usenatbib]{mnras}
\usepackage{color}
\usepackage{graphicx}
\usepackage[utf8]{inputenc}
\usepackage{amsmath, amssymb}
\usepackage{mathtools}
\usepackage{url}
\usepackage{wasysym}
\usepackage{float}
\usepackage{gensymb}
\usepackage{textcomp}
\usepackage{stmaryrd}
\usepackage{todonotes}
\usepackage{subfigure}
\usepackage{color}

\definecolor{darkgreen}{rgb}{0,0.7,0}
\title[Structure and assembly of SIDM cluster-size haloes]
{The structure and assembly history of cluster-size haloes in Self-Interacting Dark Matter}

\author[T. Brinckmann et al.]{Thejs Brinckmann$^{1,2}$\thanks{\href{mailto:brinckmann@physik.rwth-aachen.de}{brinckmann@physik.rwth-aachen.de}}, Jes\'us Zavala$^{3,2}$\thanks{\href{mailto:jzavala@hi.is}{jzavala@hi.is}}, David Rapetti$^{4,5,2}$, Steen H. Hansen$^2$,\newauthor \& Mark Vogelsberger$^6$\vspace{0.3cm}\\
$^1$Institute for Theoretical Particle Physics and Cosmology (TTK), RWTH Aachen University, D-52056 Aachen, Germany\\
$^2$Dark Cosmology Centre, Niels Bohr Institute, University of Copenhagen, Juliane Maries Vej 30, 2100 Copenhagen, Denmark \\
$^3$Center for Astrophysics and Cosmology, Science Institute, University of Iceland, Dunhagi 5, 107 Reykjavik, Iceland\\
$^4$Center for Astrophysics and Space Astronomy, Department of Astrophysical and Planetary Science, University of Colorado,\\
Boulder, C0 80309, USA\\
$^5$NASA Ames Research Center, Moffett Field, CA 94035, USA\\
$^6$Department of Physics, Kavli Institute for Astrophysics and Space Research, Massachusetts Institute of Technology,\\
77 Massachusetts Avenue, Cambridge, MA 02139, US}

\begin{document}

\maketitle

\vspace{-1in}

\begin{abstract}
We perform dark-matter-only simulations of 28 relaxed massive cluster-sized haloes for Cold Dark Matter (CDM) and Self-Interacting Dark Matter (SIDM) models, to study structural differences between the models at large radii, where the impact of baryonic physics is expected to be very limited. We find that the distributions for the radial profiles of the density, ellipsoidal axis ratios, and velocity anisotropies ($\beta$) of the haloes differ considerably between the models (at the $\sim1\sigma$ level), even at $\gtrsim10\%$ of the virial radius, if the self-scattering cross section is $\sigma/m_\chi=1$~cm$^2$~gr$^{-1}$. Direct comparison with observationally inferred density profiles disfavours SIDM for $\sigma/m_\chi=1$~cm$^2$~gr$^{-1}$, but in an intermediate radial range ($\sim3\%$ of the virial radius), where the impact of baryonic physics is uncertain. At this level of the cross section, we find a narrower $\beta$ distribution in SIDM, clearly skewed towards isotropic orbits, with no SIDM (90\% of CDM) haloes having $\beta>0.12$ at $7\%$ of the virial radius. We estimate that with an observational sample of $\sim30$ ($\sim10^{15}$~M$_\odot$) relaxed clusters, $\beta$ can potentially be used to put competitive constraints on SIDM, once observational uncertainties improve by a factor of a few. We study the suppression of the memory of halo assembly history in SIDM clusters. For $\sigma/m_\chi=1$~cm$^2$~gr$^{-1}$, we find that this happens only in the central halo regions ($\sim1/4$ of the scale radius of the halo), and only for haloes that assembled their mass within this region earlier than a formation redshift $z_f\sim2$. Otherwise, the memory of assembly remains and is reflected in ways similar to CDM, albeit with weaker trends.

\end{abstract}


\section{Introduction}
The standard model of structure formation, in which the only relevant dark matter interaction is gravitational, has been very successful in describing the large scale structure of the Universe (e.g. \citealt{Springel:2005}). It is the cornerstone of the modelling of the complexities of galaxy formation and evolution, which has made tremendous progress in the last decades as reflected by the current generation of hydrodinamical simulations in a full comological setting (see e.g. \citealt{2014MNRAS.444.1518V,2014Natur.509..177V,Dubois:2014,Schaye:2014tpa}). Nevertheless, the study of non-gravitational dark matter interactions in structure formation has become more relevant in the last few years thanks to the appeal of finding a dynamical signature of new dark matter physics in the properties of galaxies. In particular, the possibility of dark matter being strongly self-interacting (SIDM) is an attractive possibility from the point of view of both astrophysics, and the particle physics of dark matter (e.g. \citealt{Spergel:1999mh,Yoshida:2000uw,Gnedin:2000ea,Firmani:2000qe,Dave:2001,Colin:2002nk,ArkaniHamed:2008qn,Ackerman:mha,Feng:2009mn,Buckley:2009in,Feng:2009hw,Loeb:2010gj,Aarssen:2012fx,Tulin:2012re,Vogelsberger:2012ku,Rocha:2012jg,Peter:2012jh,Zavala:2012us,Robertson:2015faa,Cyr-Racine:2016,Vogelsberger:2015gpr,Harvey:2016bqd,Kamada:2016euw}). In SIDM, dark matter particles can collide with themselves with a cross section of similar amplitude to that of the strong force. If the new interactions governing dark matter particles are mostly hidden (i.e. they do not result in standard model byproducts), then the only hope to detect such interactions is through their imprint in the formation and assembly history of dark matter structures.

Structure formation in a SIDM Universe has been studied extensively by now, mostly in settings where the dynamical impact of the luminous matter is neglected, and with the goal of constraining the amplitude of the self-interacting transfer cross section per unit mass $\sigma/m_\chi$. On the lower end, it is probably unlikely that signatures of SIDM can be identified by dynamical studies of the poperties of galaxies if  $\sigma/m_\chi<0.1$~ cm$^2$~gr$^{-1}$ (see e.g. \citealt{Zavala:2012us}). On the higher end, the current most stringent constraints on the cross section are at around $1-2$~ cm$^2$~gr$^{-1}$ (e.g. \citealt{Peter:2012jh,Harvey:2015,Robertson:2017,Wittman:2017}). The debate over the robustness of these upper constraints in the cross section remains open, due to (i) the difficulty of properly measuring observables that constrain the cross section (e.g. the offsets between galaxies and dark matter in merging clusters) and (ii) the difficulty in interpreting these observables given the uncertain impact of the gas and stellar physics that create the luminous galaxies (henceforth baryonic physics), and its synergy with the SIDM physics \citep{Kaplinghat:2016,Elbert:2016,Robles2017}. 

Despite these difficulties, it is important to remark that there is a relatively narrow window for a {\it constant} cross section SIDM model ($0.1$~cm$^2$~gr$^{-1}<\sigma/m_\chi\lesssim$ 1~cm$^2$~gr$^{-1}$) to be a viable and interesting possibility for structure formation. Efforts to close this window (or find signatures of SIDM in the process) are therefore timely and necessary. In this work, we perform a detailed analysis of the structure of {\it relaxed} massive cluster-size haloes using a suite of $N$-body simulations covering the extreme cross section values within this window. Although our simulations do not include baryonic physics, we concentrate most of our analysis in the outer enough regions of galaxy clusters where the SIDM effects are still significant, while the impact of baryonic physics is less relevant than in the central regions. 

In particular, we concentrate on $10^{15}$~M$_\odot$ {\it dynamically relaxed} haloes and study three main structural radial profiles: density, shape and velocity anisotropy. In addition to quantifying the structural differences between constant cross section SIDM haloes and CDM, we explore whether an imprint of the assembly history of haloes remains in the structural properties of SIDM haloes at $z=0$, as it does in collisionless CDM haloes.  

The paper is organised as follows. In Section \ref{sec:sims} we introduce our simulations and sample of haloes. In Section 3 we present our results: in \ref{sec:relaxation} we look at the relaxation properties of our haloes, while in \ref{sec:density}$-$\ref{sec:beta} we examine the density, shape, and velocity anisotropy profile distributions of haloes across the different models. In Section \ref{sec:assembly} we discuss the memory of the assembly history that remains in SIDM haloes. Finally, in section \ref{sec:conclusions} we give a summary and present our conclusions.
\section{Numerical simulations}
\label{sec:sims}
Our $N$-body simulations were run using the \texttt{AREPO} code~\citep{Springel:2009aa}, with an extra module for computing dark matter self-interactions (\citealt*{Vogelsberger:2012ku}; \citealt{Vogelsberger:2015gpr}). This module computes the elastic scattering between pairs of particles via an $N$-body/Monte Carlo method, which simulates the macroscopical consequences of the interactions in a probabilistic manner, with the local interaction rate given by
\begin{align}
\Gamma_{\rm loc} = \rho_{\rm loc} \left(\frac{\sigma}{m_{\chi}}\right )v_{\rm rms,loc}
\end{align}
where $\rho_{\rm loc}$ is the local density and $v_{\rm rms,loc}$ is the local root mean square velocity. The details and testing of the algorithm for collisions can be found in \citet{Vogelsberger:2012ku}. In this work we only study the case of a constant scattering cross section, and in particular we perform simulations for two cases $\sigma/m_{\chi} =$~1~cm$^2$~gr$^{-1}$ (SIDM1) and $\sigma/m_{\chi} =$~0.1~cm$^2$~gr$^{-1}$ (SIDM0.1), in addition to the vanilla CDM case for comparison. Our simulations use cosmological parameters consistent with a Planck cosmology: $\Omega_m=0.315$, $\Omega_\Lambda=0.685$, $h=0.673$, $\sigma_8=0.83$, and $n_s=0.96$, where $\Omega_m$ and $\Omega_\Lambda$ are the contributions from matter and cosmological constant to the mass/energy density of the Universe, respectively, $h$ is the dimensionless Hubble constant parameter at redshift zero,
$n_s$ is the spectral index of the primordial power spectrum, and $\sigma_8$ is the rms amplitude of linear mass fluctuations in $8$~Mpc~h$^{-1}$
spheres today, at $z=0$. 

\begin{figure*}
	\centering
	\qquad
	\includegraphics[width=0.40\linewidth]{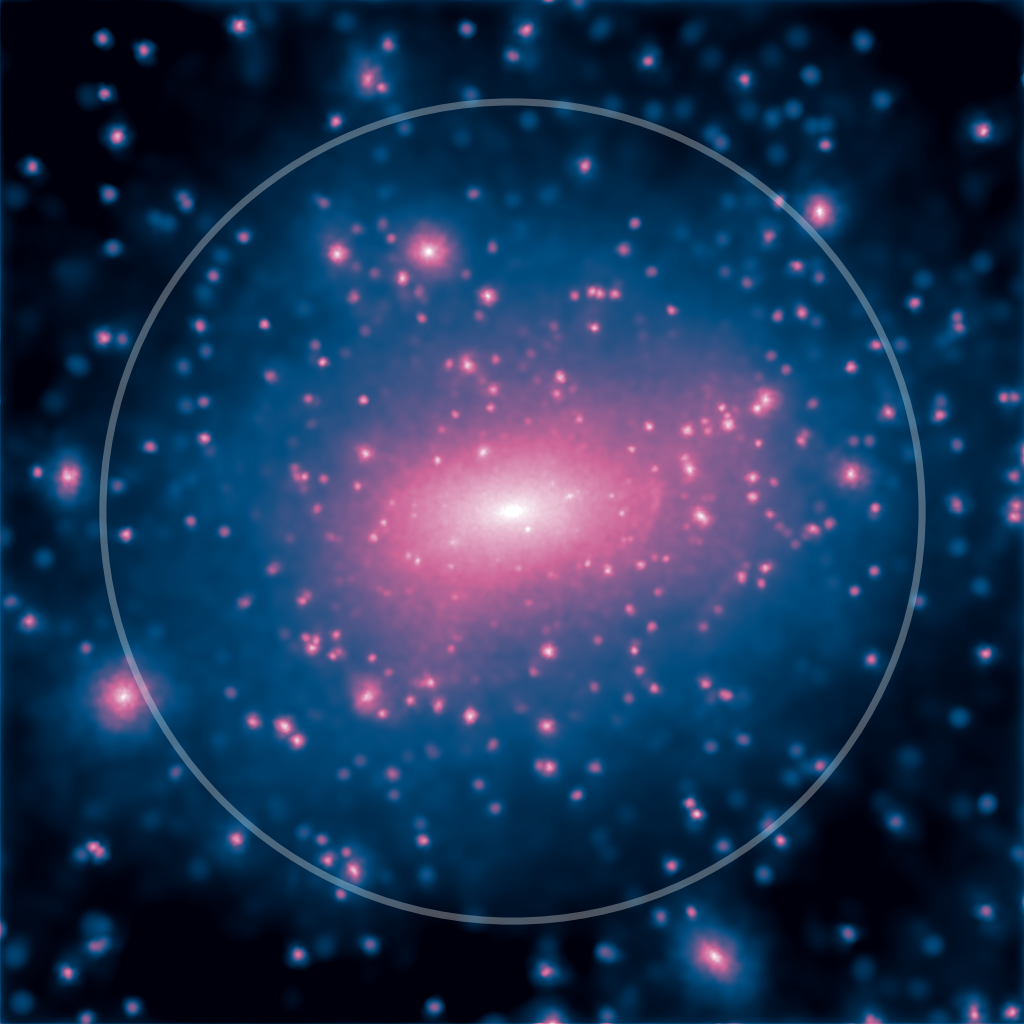}\hfill
	\includegraphics[width=0.40\linewidth]{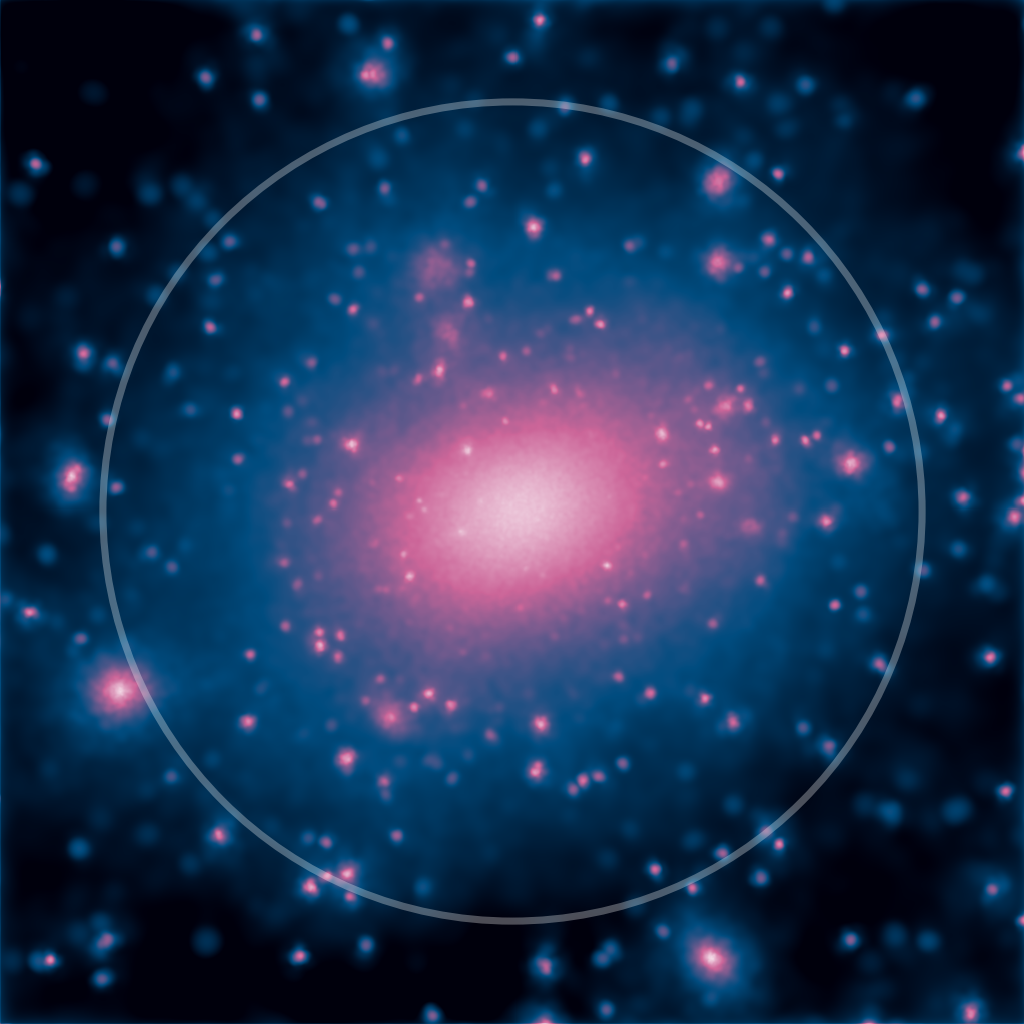}
	\qquad
	\caption{The most massive halo in our sample (M$_{200}\sim2\times10^{15}$~M$_\odot$ h$^{-1}$) in the CDM (left) and SIDM1 (right) cases. The circle marks the virial radius of the halo ($R_{200}\sim2$~Mpc h$^{-1}$).}
	\label{fig:sims:vis}
\end{figure*}

Our analysis is based on zoom resimulations of a sample of 28 cluster-sized haloes sampled from a large 1~Gpc$^3$~h$^{-1}$ parent simulation with $512^3$ dark matter particles. The sample is chosen with the aim of studying dynamically relaxed systems, which could potentially be compared with observed clusters with no obvious sign of recent major mergers. With this purpose in mind, we adopt the criteria described in \citet{Ludlow:2013bd} to select a sample of relaxed haloes from our parent simulation:\\ 
\textbf{(i)} The subhalo mass fraction must be below 10\% (f$_{\rm sub}$~=~M$_{\rm sub}$/M$_{200}<$~0.1)\footnote{Throughout this work we define virial masses and radii according to the radius enclosing an average density of 200 times the critical density of the Universe.}. Otherwise, very massive subhaloes might perturb the halo significantly.\\
\textbf{(ii)} The distance between the center of mass of the halo and its potential minimum, relative to the virial radius, must be less than 0.07 (d$_{\rm off}$~=~$\vert$r$_{\rm pot}$~$-$~r$_{CM}\vert/R_{200}<$~0.07). A larger difference indicates a substantial merger is in progress.\\
\textbf{(iii)} The `virial ratio' of total kinetic to potential energies in a given halo must be less than 1.35 ($2T/\vert U\vert<$~1.35), which is a threshold to separate cases with stronger departures from virialization, i.e. dynamical equilibrium.\\

The zoom simulations have an effective resolution of $4096^3$ particles in the highest resolution region, which is surrounded by regions of intermediate resolution and finally a low resolution volume with an effective resolution of $256^3$ particles. To construct the initial conditions of the zoom simulations we followed closely the methodology described in e.g. \citet{Onorbe:2013fpa}:
\begin{itemize}
\item Pick the sample of 28 most massive ``relaxed'' haloes in the parent simulation, as described above.
\item Select the Lagrangian region around each of these haloes at $z=0$ in the parent simulation. This is the target region for resimulation.
\item Traceback the particles to the initial target redshift for resimulation ($z=50$) by matching the unique particle ID numbers across redshifts.
\item Compute the initial conditions for the zoom simulation using the code \texttt{MUSIC}\footnote{https://people.phys.ethz.ch/$\sim$hahn/MUSIC/} \citep{Hahn:2011uy}, specifying the ellipsoidal (or cuboid) region containing the targeted particles at $z=50$ as the high resolution region {(see Appendix \ref{sec:convergence} for more details and convergence tests)}.
\end{itemize}

For the high resolution region, the effective Plummer equivalent gravitational softening length is $\epsilon = 5.4$~kpc~h$^{-1}$, while the particle mass is m$_p = 1.271 \times 10^9$~M$_{\astrosun}$~h$^{-1}$.

Our final simulation suite consists of 28 haloes simulated with the same initial conditions in CDM, SIDM1 and SIDM0.1, with a virial mass and radius range in between: R$_{200} \approx 1300-2000$~kpc~h$^{-1}$, and M$_{200} \approx 0.5-1.9 \times 10^{15}$~M$_{\astrosun}$~h$^{-1}$. Except for the most massive cluster, the sample has a narrow distribution centered around M$_{200} \sim 0.9 \times 10^{15}$~M$_{\astrosun}$~h$^{-1}$ and R$_{200} \sim 1550$~kpc~h$^{-1}$ (see figure \ref{fig:sims:info}). A visual impression structural differences between CDM and SIDM haloes is given in Figure \ref{fig:sims:vis}, where we show dark matter density projections for the
most massive of our haloes for CDM and SIDM1 in the left and right panels, respectively.
For each simulation, we have created halo catalogues, first by using the friends-of-friends (FOF) algorithm and then using the \texttt{SUBFIND} algorithm \citep{Springel:2000qu} to identify selfbound (sub)haloes. The particles within the main halo of a given structure are the main focus of our study.

We note that for the main halo properties analysed in this work -- density, halo shape, and velocity anisotropy radial profiles -- we performed convergence tests to determine the spatial resolutions we can trust. These are described in Appendix A.

\section{Results}
\subsection{Relaxation}
\label{sec:relaxation}
Having defined our halo relaxation criteria in section \ref{sec:sims}, we now study how our ensemble of haloes differ between the CDM and SIDM1 parent simulations in regards to their equilibrium states (there is a negligible difference between CDM and SIDM0.1) by looking at all haloes with more than 500 particles. We find that the number of haloes satisfying our relaxation criteria differ significantly between the two cosmologies, with almost 20\% more relaxed haloes in SIDM1 at $z=0$ (40\% if we only examine the most massive haloes with more than 1000 particles, see Table \ref{tab:relaxation}).

Examining each criteria separately, we find that the virialization threshold, $2T/\vert U\vert<$~1.35, is the most important one in explaining this difference (this holds up to $z\sim1$; the number of resolved haloes drops quickly above this redshift). The median of the distribution of $2T/\vert U\vert$ values is approximately 0.5$-$1\% lower in SIDM1 than in CDM ($0<z<1$). We interpret this result as a consequence of the inside-out `heat' transfer that occurs during dark matter self-interactions, which leads to the thermalization of the central regions. Despite commonly assumed to impact only the innermost regions of haloes, we find that self-interactions with a cross section of 1~cm$^2$~gr$^{-1}$ are strong enough to affect the global virial ratio of the entire halo.

\citet{Kim:2016ujt} found that dark matter self-interactions ultimately shorten the timescales of halo mergers, despite competition between the enhanced momentum exchange of colliding haloes and a stalling of dynamical friction in nearly constant density cores. From this we would expect fewer on-going mergers and therefore more relaxed SIDM halos compared to CDM, as indeed we have found here.

\begin{table}
\caption{The number of relaxed SIDM1 haloes divided by the number of relaxed CDM haloes, N$_{\rm relax, SIDM1}$/N$_{\rm relax, CDM}$, for redshifts 0 to 0.5. Column two shows the ratio for haloes with more than 500 particles and column three for haloes with more than 1000 particles. The statistics are increasingly poor with greater redshift for the most massive haloes, since there are not enough haloes fulfilling the exclusion criteria. For the less massive haloes there are $\sim 20\%$ more relaxed SIDM haloes, whereas for the more massive haloes the ratio is much larger (although the number count of objects is very low).}
\label{tab:relaxation}
\centering
\begin{tabular}{c c c}
z & N$_{\rm par}>500$ & N$_{\rm par}>1000$ \\ \hline
0 & 1.17 & 1.4 \\
0.1 & 1.33 & 1.5 \\
0.2 & 1.17 & 1.5 \\
0.3 & 1.24 & 1.6 \\
0.4 & 1.18 & - \\
0.5 & 1.17 & - 
\end{tabular}
\end{table}

The haloes used to define the sample we resimulate all contain more than 1000 particles in the parent simulation. This is a relatively low particle number threshold that could in principle artificially bias the selected sample toward larger values of the spin parameter and the virial ratio ($2T/|U|$), due to poor particle sampling (e.g. \citealt{Bett:2006zy}). However, we note a relevant mitigating circumstance. By imposing a lower limit in the virial ratio in our relaxation criteria, we are also avoiding the most problematic haloes which are affected by poor particle sampling. \citet{Bett:2006zy} pointed out that choosing a lower limit mitigates the issue, and that the impact on the median distribution is negligible for haloes with more than $\sim300$ particles. Recently, \citet{2017MNRAS.471.2871B} revisited this issue and found that even though the median spin values might be fine, the error on the spin of individual haloes can be substantial ($\sim$100\%) for haloes sampled with $\sim300$ particles. However, near 1000 particles, which is our case, the situation improves substantially, with individual errors below 50~\%, and no bias in the median values. This make us confident that the halo sample we have selected is not strongly biased due to poor particle sampling in the parent simulation. We note that in any case, since our main interest is that of a comparison between CDM and SIDM based on equal initial conditions, the conclusions that follow from our results are not affected by poor particle sampling, since they are based on re-simulations that are not affected by this effect.

In the following, we concentrate on three main structural characteristics of dark matter haloes: (i) spherically averaged density profiles; (ii) departures from spherical symmetry (halo shapes) and (iii) departures from isotropy in the velocity dispersion tensor (velocity anisotropy).

\subsection{Density profiles}
\label{sec:density}
The centre of each halo is given by the position of the most bound particle (i.e., the minimum of the gravitational potential of all bound particles).
From this centre, we compute the spherically averaged radial density profile.
Due to limited resolution, we can only trust the density profile down to certain radius. For the most massive halo in our CDM simulation (M$_{200} = 1.95 \times 10^{15}$~M$_{\astrosun}$~h$^{-1}$ and R$_{200} = 2030$~kpc~h$^{-1}$) we did a convergence test with three levels of resolution: our reference level with $\epsilon=5.4$~kpc~h$^{-1}$, one with half the resolution ($2\epsilon$) and one with twice the resolution ($\epsilon/2$). For CDM, the test is consistent with what is commonly used in the literature as a radius of convergence for the density profile: $\sim4\epsilon$, with about $4\%$ difference between consecutive resolution levels at this radius (see Appendix \ref{sec:convergence}).
The SIDM simulations typically have better convergence since once the SIDM core is apparent, a relatively low resolution level
can reliably reveal the central core density and the extent of the core, whereas in CDM, the density is ever increasing at lower
radii (e.g. \citealt{Vogelsberger:2012ku,Elbert2015}). We take $4 \epsilon$ as a conservative choice for the minimum radius we can trust for the density profiles.

\begin{figure}
\centering
\includegraphics[width=\linewidth]{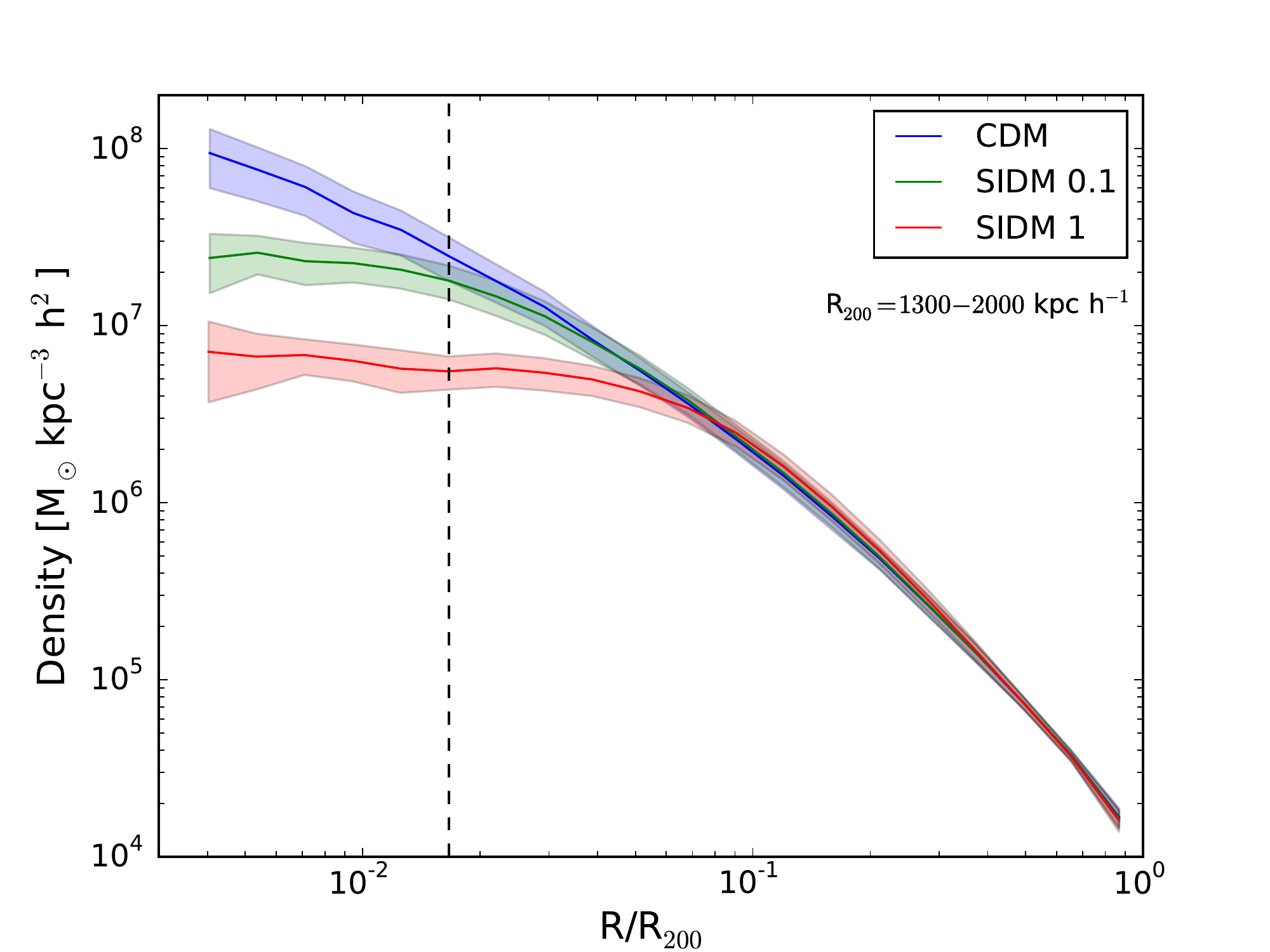}
\caption{Distribution of radial density profiles for a sample of 28 cluster-sized haloes in CDM (blue), SIDM0.1 (green) and SIDM1 (red). The radius of each halo is scaled by its virial radius (R/R$_{200}$). The median of each distribution is shown as a solid line and the shaded region indicates the standard deviation of the distribution. The dashed line indicates $4\epsilon/R_{200, min}$, where R$_{200,min}$ is the virial radius of the smallest halo in our sample. This line is a reference to the radius we can reliably trust in our combined simulation sample. While the SIDM0.1 case shows small departures from CDM, less than a 1$\sigma$ separation between the distributions at the trust radius, the SIDM1 case shows larger extended cored haloes with a clear separation already at 0.05~R/R$_{200}$ ($\sim 75-100$)~kpc~h$^{-1}$.}
\label{fig:results:density}
\end{figure}

Fig.~\ref{fig:results:density} shows the distributions of density profiles for all 28 haloes for our three models, CDM (blue), SIDM1 (red) and SIDM0.1 (green).
The solid lines are the medians of each distribution, while the shaded regions indicate the $\pm 1 \sigma$ regions of each distribution. Since we are combining all haloes into the same plot, we use the smallest halo (R$_{200}=1300$~kpc~h$^{-1}$) to establish the minimum dimensionless radius we can trust in our simulations (4$\epsilon$/R$_{200} \sim 1.5 \times 10^{-2}$, dashed line). Fig.~\ref{fig:results:density} clearly shows important differences between CDM and the SIDM simulations in the resolved regime of our simulations. The SIDM1 haloes have an extended core with a size of R$_{\rm core}/R_{200} \sim 6 \times 10^{-2}$ and a central density of $\sim 6 \times 10^6$~M$_{\astrosun}$~kpc$^{-3}$~h$^2$, a factor of $\sim 4$ smaller than CDM at the trust radius of $4\epsilon$/R$_{200} \sim 1.5 \times 10^{-2}$. The SIDM0.1 case is much closer to CDM, but our resolution is enough to clearly separate the halo distributions (albeit at less than 1$\sigma$ level) at $4\epsilon$/R$_{200}$. This is a radius well within the inner region of the clusters ($\sim 20$~kpc~h$^{-1}$), where the effects of the luminous matter on the dark matter distribution are quite important. This makes the cross section level of 0.1~cm$^2$~gr$^{-1}$ a difficult case to obtain conclusions based on simulations without baryonic physics. For the SIDM haloes with $\sigma/m_\chi=$~1.0~cm$^2$~gr$^{-1}$, the situation is more promising, since in this case there is a clear separation already at R/R$_{200} \sim 0.05$ ($\sim 75-100$~kpc~h$^{-1}$), where the influence of baryons is less relevant.

\begin{figure}
	\centering
	\includegraphics[width=\linewidth]{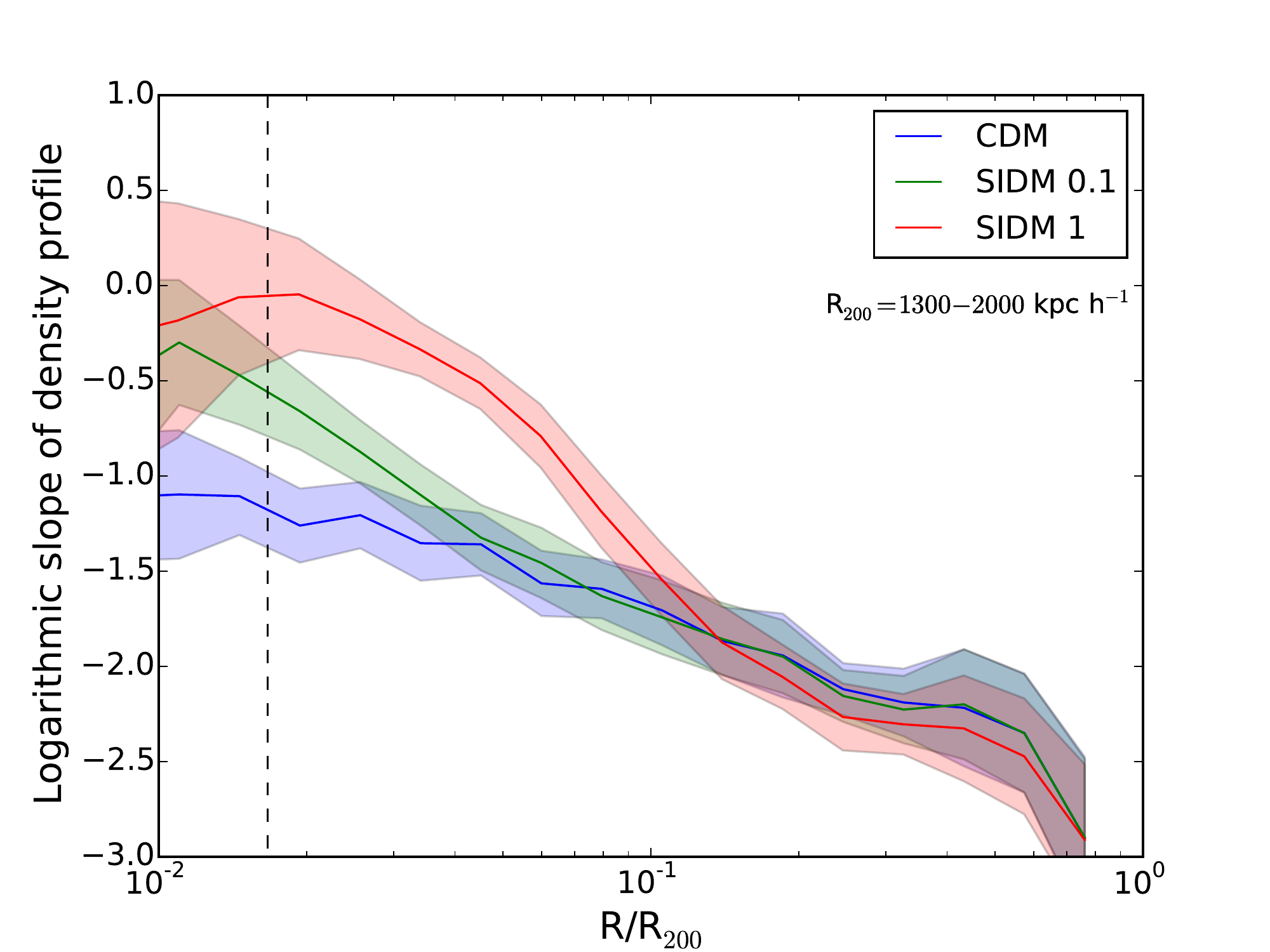}
	\caption{Distribution of the logarithmic slope of the radial density profiles for our cluster-sized halo sample in CDM (blue), SIDM0.1 (green) and SIDM1 (red). For each halo the radius is scaled by its virial radius (R/R$_{200}$). The solid lines indicate the median of each distribution and the shaded regions show the standard deviation of the distributions. The dashed line is $4\epsilon/R_{200, min}$, indicating the radius to which we can reliably trust our combined simulation sample. We see that the difference between CDM and SIDM is significantly enhanced compared to the one seen in the comparison between density profiles (Fig.~\ref{fig:results:density}), with $1\sigma$ separation at $\sim 40-50$ ($\sim 125-150$)~kpc~h$^{-1}$ for SIDM0.1 (SIDM1).}
	\label{fig:results:slope}
\end{figure}

In Fig. \ref{fig:results:slope} we show the logarithmic slope of the density profile $\gamma$ as a function of radius. At small radii, this figure shows the expected cored-like and NFW-like ($\gamma\sim-1$) behaviour for CDM and SIDM1, respectively. At larger radii, the difference in the profiles is enhanced. For instance, the distribution of slopes for SIDM 0.1 differs from CDM at the $1\sigma$ level at
$\sim40-50$~kpc~h$^{-1}$ ($\sim 2.5 \times 10^{-2}$~R/R$_{200}$), which is a larger radius than the analogous for the density profile plot ($4\epsilon/$R$_{200}\sim 1.5 \times 10^{-2}$). Although the difference seen here is significant, we note that baryonic effects are expected to be more relevant than this at $\sim 50$ kpc, making it more difficult to draw firm conclusions without the added effects of baryonic physics in the simulations. For the case of SIDM1, we see significant departures from CDM already at R/R$_{200}\sim0.1$. We note also that due to the redistribution of mass from the inside-out in SIDM, the density profile beyond R/R$_{200}\sim0.1$ is actually steeper than CDM out to radii close to the virial radius. Although the difference is small, Fig.~\ref{fig:results:slope} shows a visible separation of the distributions between CDM and SIDM1 in the logarithmic slope in these outer regions.

From the cosmological hydrodynamical simulations of the EAGLE project, \citet{Schaller:2015} found that the density profile of EAGLE haloes present only minimal differences beyond R/R$_{200}=0.05$, relative to the simulations without baryonic physics. A similar conclusion, albeit at slightly larger radii was found in the Horizon-AGN simulation project \citep{Peirani:2016}. Although the  mass range of these simulations does not reach clusters as massive as the ones studied here, their results suggest that baryonic physics should not substantially modify the density profile at the radius at which a cross section of 1~cm$^2$~gr$^{-1}$ in SIDM does. Instead, the effects of an order of magnitude smaller cross section in SIDM are relevant
within the region where baryonic physics is important. Whether the inclusion of baryonic physics makes the dark matter halo contract or expand in the central regions, depends on the competition between the opposite effects of gas cooling and AGN/SNe feedback. If feedback is sufficiently strong it can create a cored-like distribution similar to the one produced in SIDM. A careful study of the synergy between the dark and baryonic processes would be needed to obtain realistic predictions of SIDM in the central regions. However, as we have argued, at radii near $10\%$ of the virial radius, the impact of baryonic physics is expected to be small enough to be able to neglect it when using $\gamma$ to constrain the SIDM cross section at the higher end of the current level of interest ($0.1<\sigma/m_x<1$~cm$^2$~gr$^{-1}$).

\begin{figure}
	\centering
	\includegraphics[width=\linewidth]{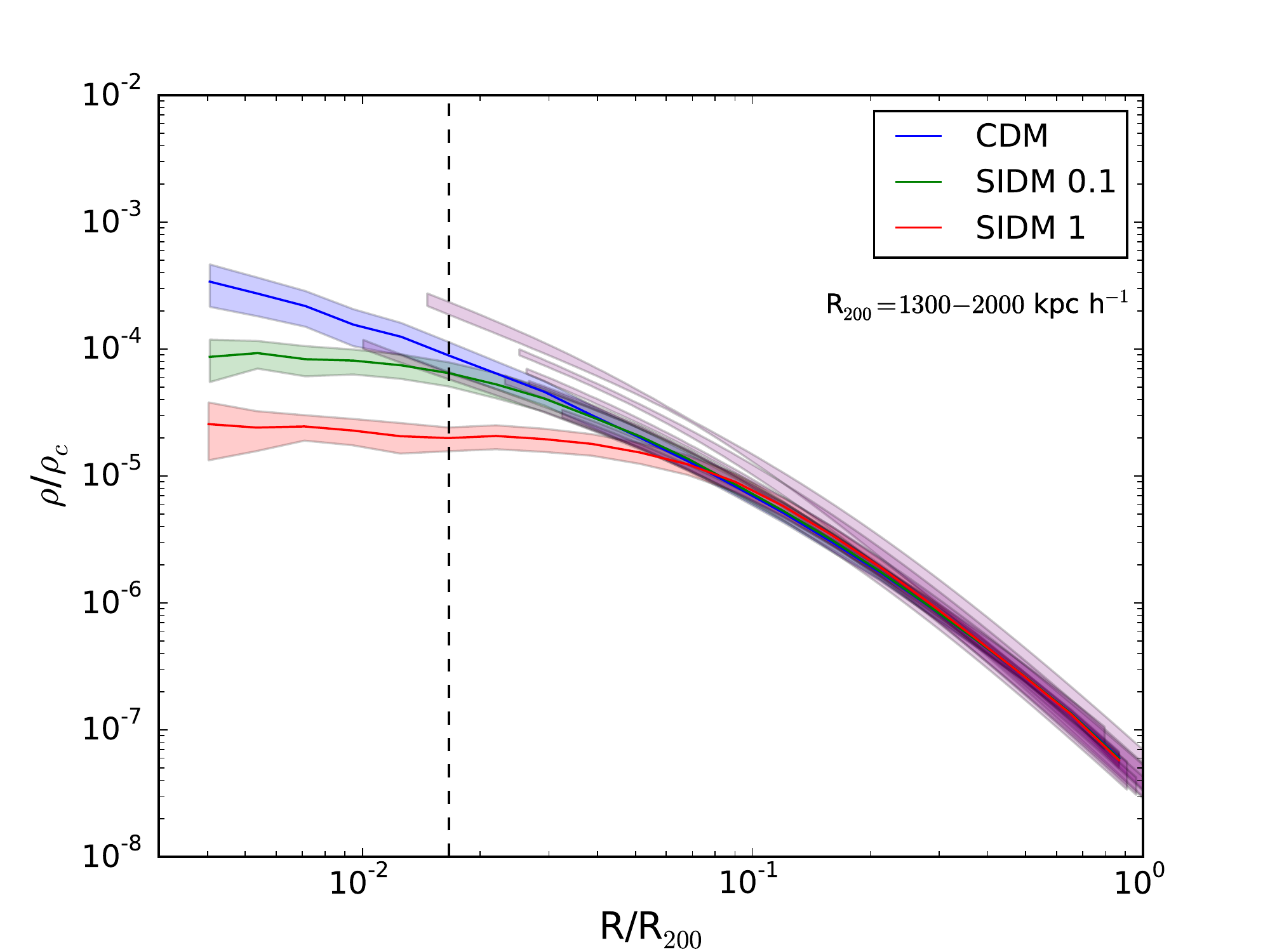}
	\caption{Comparison of the distribution of radial density profiles from our simulation as in Fig. \ref{fig:results:density} with the inferred dark matter density profiles for seven galaxy clusters from \citet{Newman:2012nw} in purple. Densities are rescaled by the critical density at the correponding redshift for each case ($z=0$ for our simulations) and the observed redshift for the different galaxy clusters. The radial range of the observed profiles is bounded at low radii by the half light radius of the respective Brightest Central Galaxy, while the width is given by the $1\sigma$ random and systematic errors. This comparison seemingly disfavours SIDM1, but the impact of baryonic physics in the SIDM distribution, absent in our simulations, cannot be neglected in the radial range where the disagreement is strong.}
	\label{fig:results:density_newman}
\end{figure}

We show a direct comparison between the density profiles of our simulations with the observationally inferred dark matter density profiles of seven galaxy clusters from \citet{Newman:2012nw} in Fig. \ref{fig:results:density_newman}. The densities are scaled by the critical density at the redshift of the halo or galaxy cluster. For each of the galaxy clusters, the profiles have a cutoff at low radii given by the half light radius of their respective Brightest Cluster Galaxy (BCG), which gives an indication of the boundary where the impact of the baryonic distribution is clearly dynamically important. The width of the observational profiles are given by the $1\sigma$ random and systematic errors. From Fig.~\ref{fig:results:density_newman} we see a strong indication that SIDM1 is not easily reconciled with observed density profiles. We nevertheless emphasize that the impact of baryons on a SIDM halo, even at the scales beyond which the BCG is relevant, is non-trivial and must be studied with care, particularly the impact on the evolution of a SIDM halo by the assembly of the cluster. As we mentioned above, simulations within the CDM cosmology show that an impact of baryonic physics in the dark matter density profile is to be expected up to $0.05$~R$_{200}$, which is the scale at which observations start to depart strongly from the SIDM1 case. We thus refrain from giving a quantitative constraint on SIDM based on these observed profiles and leave this for a future work, where the interplay between SIDM and baryonic physics is taken into full consideration.

\subsection{Halo shapes}
\label{sec:shapes}
To study the halo shapes of the simulated haloes, we developed our own code based on the methodology described in \citet{Zemp:2011ed}. The code\footnote{https://github.com/brinckmann/halo\_shapes} computes the eigenvectors and eigenvalues of the shape tensor of a distribution of particles, in order to determine the orientation and magnitude of the principal axes of the distribution. For our analysis, we divide each halo into a number of ellipsoidal shells, and for each shell our code iteratively finds the shape of the shell by calculating the eigenvectors and eigenvalues of the shape tensor of the particles within the shell, until convergence is achieved (when the axis ratio of both the minor and intermediate axis to the major axis is deviating by less than 0.1\% over the last 10 iterations). It typically takes between 20 to 40 iterations to reach convergence, but a low number of particles or a very spherical system (where the direction of the eigenvectors becomes ambiguous) can cause the code to fail to converge. For each iteration, the code deforms the volume of the ellipsoidal shell according to the previous result (adding or removing particles in the process), while keeping the size of the major axis constant. This means a new set of particles is selected for each iteration, until convergence is obtained and the set of particles is roughly constant.

The shape tensor is defined as
\begin{align}
\boldsymbol{S} \equiv \frac{\boldsymbol{M}}{M_{\rm tot}} = \frac{\int_V \rho(\boldsymbol{r})\boldsymbol{r}\boldsymbol{r}^T \rm d V}{\int_V \rho(\boldsymbol{r}) \rm d V},
\end{align}
where $\rho$ is the density and $\boldsymbol{r}$ is the position vector relative to the center of the mass distribution. $M_{\rm tot} = \int_V \rho(\boldsymbol{r}) \rm d V$ is the total mass of the system and $\boldsymbol{M}$ is the second moment of the mass distribution, which is related to the moment of inertia tensor via
\begin{align}
\boldsymbol{I} = \rm tr(\boldsymbol{M})\boldsymbol{1} - \boldsymbol{M},
\end{align}
where $\boldsymbol{1}$ is the identity tensor. Discretizing the individual elements of the shape tensor we have
\begin{align}
S_{ij} = \frac{\sum_k m_k (\boldsymbol{r}_k)_i (\boldsymbol{r}_k)_j}{\sum_k m_k},
\end{align}
where $m_k$ is the mass of the k-th particle and $(\boldsymbol{r}_k)_i$ is the i-th component of the position vector of the k-th particle. Since in our case all the particles have the same mass, we are left with
\begin{align}
S_{ij} = \frac{\sum_k  (\boldsymbol{r}_k)_i (\boldsymbol{r}_k)_j}{N_p},
\end{align}
where $N_p$ is the number of particles in the shell.

In addition to the default option that we use for our main analysis, which keeps the magnitude of the major axis of each shell fixed in the iteration procedure, the code can instead keep the volume of the shells constant. There is also the option to calculate the shapes based on the entire enclosed mass within a given ellipsoid (as opposed to the mass within the ellipsoidal shell). The contribution of each particle to the shape tensor can also be weighted by the elliptical radius $r_{\rm ell} = \sqrt{x_{\rm ell}^2 + y_{\rm ell}^2 \left(b/a\right)^{-2} + z_{\rm ell}^2 \left(c/a\right)^{-2}}$, where $x_{\rm ell}$, $y_{\rm ell}$ and $z_{\rm ell}$ are the distances along the principal axes of the ellipsoid and $a, b, c$ are the major, intermediate and minor axes, respectively. In Appendix \ref{sec:code_testing}, we present the performance of our code through tests with predefined halo shape configurations, with variations over the options described above. For the purposes of our comparison between CDM and SIDM haloes, we find that the default option is suitable enough.   

To have an indication of the minimum radius we can trust in our halo shape profiles we use the Power radius \citep{Power:2002sw}. Following \citet{Springel:2008cc}, this is given by the minimum radius that satisfies
\begin{align}
\frac{\sqrt{200}}{8} \frac{N(r)}{\ln \left[N(r)\right]} \left[\frac{\bar{\rho}(r)}{\rho_{\rm crit}} \right]^{-1/2} \ge 1,
\end{align}
where $N(r)$ is the number of particles within a radius r, $\rho_{\rm crit} = 3 H_0^2 / (8 \pi G)$ is the critical density of the Universe today, and $\bar{\rho}(r)$ is the mean density within $r$
\begin{align}
\bar{\rho}(r) = \frac{N(r) m_p}{r^3},
\end{align}
where $m_p$ is the particle mass. For an NFW profile with an enclosed mass profile $M_{\rm NFW}(r)$ \citep*{Navarro:1996gj}, which is a good description of CDM haloes, we can write $N(r) = M_{\rm NFW}(r)/m_p$.
Using the values of $V_{\rm max}$, which is the maximum circular velocity, and $r_{\rm max}$, the radius where $V_{\rm max}$ is attained, from each halo as provided by \texttt{SUBFIND}, we can compute the NFW mass profile for each halo and estimate the Power radius. For the haloes in our sample, this is approximately r$_{\rm Power} \approx 40$~kpc~h$^{-1}$, at the resolution level we use for our main analysis. For our sample of haloes we adopt as our scaled trust radius the ratio r$_{\rm Power}$/R$_{200}$ for the smallest halo, as the most conservative choice (in Appendix \ref{sec:code_testing} we show that our halo shapes code can be trusted to near this radius for CDM; we discuss the SIDM1 case later).

\begin{figure}
\centering
\includegraphics[width=\linewidth]{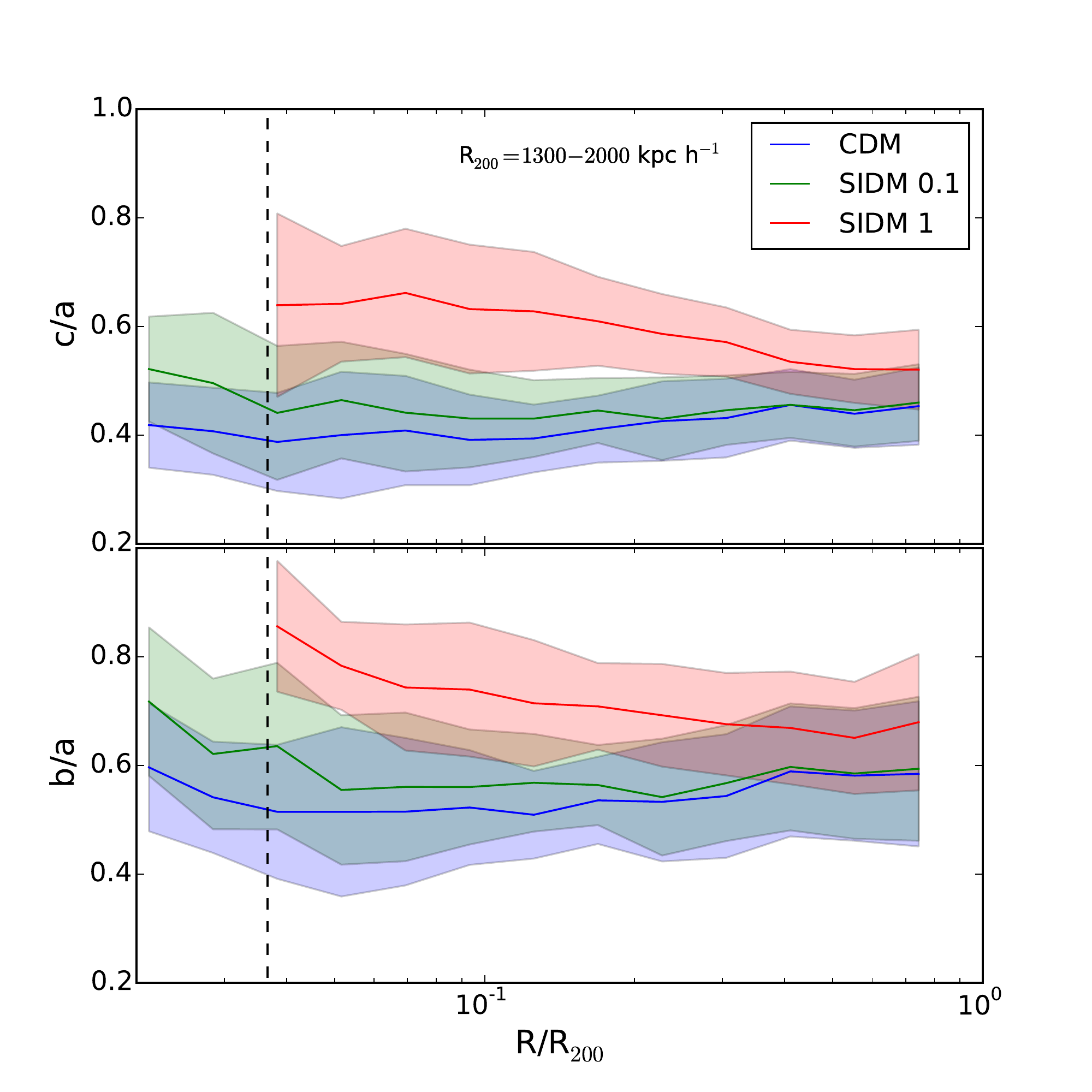}
\caption{\textbf{Top}: Ratio of the minor to major axis (c/a) vs radius scaled by the virial radius of each halo (R/R$_{200}$). \textbf{Bottom}: Similar to the top panel, but for the intermediate to major axis (b/a). These results correspond to the combined distributions of 28 haloes for each of the three different simulation types (CDM in blue, SIDM0.1 in green and SIDM1 in red). The central lines are given by the medians of the distributions and the shaded regions by the standard deviations. The trust radius, r$_{Power}$/R$_{200,\rm min}$ (see section \ref{sec:shapes}), is indicated as a vertical dashed line. We see that for SIDM1, the trust radius is too optimistic. This is due to the presence of a core in these haloes (see Figure \ref{fig:results:density}), resulting in difficulties for the halo shapes code to reliably converge (see text for discussion). Nevertheless, the differences between SIDM1 and CDM are clear at all radii. We see a hint of divergence between CDM and SIDM0.1 at R/R$_{200} < 0.1$, but the statistics are too poor to draw any firm conclusions.
}
\label{fig:results:shapes_ca}
\end{figure}

Fig. \ref{fig:results:shapes_ca} shows the combined halo shape profile distributions for our halo sample as given by the radial dependence of the principal axis ratios $c/a$ and $b/a$, top and bottom panels, respectively. The solid lines indicate the medians of the distributions, while the shaded regions indicate the $\pm 1 \sigma$ regions. We find a clear separation (at the $\sim 1 \sigma$ level) between CDM and SIDM1 in cluster-sized haloes across a large radial range, $5 \times 10^{-2} \lesssim$~R/R$_{200} \lesssim 3 \times 10^{-1}$ ($\sim 100-500$~kpc~h$^{-1}$). At larger radii the differences get progressively smaller, while in the inner regions the problems are related to resolution and difficulties in obtaining reliable shapes for cored distributions (see Appendix \ref{sec:code_testing}). Remarkably, the distributions of axis ratios between both cosmologies can be separated distinctively (at the $\sim 1 \sigma$ level), at considerably larger radii than the distributions of density profiles. This potentially makes halo shapes a diagnostic to test dark matter collisionality that is less sensitive to baryonic effects. 

We note that although our halo shapes code has some biases (see Appendix \ref{sec:code_testing}), these are expected to be of similar magnitude for both CDM and SIDM1 (in the regions that are free from resolution issues), and considerably smaller than the difference we see in Fig. \ref{fig:results:shapes_ca} related to the nature of dark matter.
The SIDM0.1 case on the other hand, is too close to CDM across all relevant radii. To study the way the distributions of halo shapes separate at a small fixed radius we would need to significantly increase our simulated halo sample.

\citet{Peter:2012jh} made a detailed study of halo shapes in SIDM simulations, using the two values of constant cross section we have used here. They performed simulations of a single resolution in a cosmological box, instead of zoom simulations as we have performed for this work. Given the maximum size of their simulated volume, they did not probe the massive cluster regime we are studying here. They obtained distributions of halo shapes in CDM and SIDM for 50 haloes in the mass range $10^{13}-10^{14}$~M$_\odot$~h$^{-1}$ (see the right panel of their Fig. 3). Their results are qualitatively similar to ours beyond $10\%$ of the virial radius, while in the inner regions they obtained a larger separation between CDM and SIDM1, with the latter leading more rapidly towards spherical shapes at smaller radius. This could be the result of different biases in the codes used to compute the halo shapes. From the study of \cite{Zemp:2011ed}, which we confirm with our analysis, computing the halo shapes is particularly difficult in cored density profiles that get monotonically more spherical towards the centre. In this case, there is a substantial bias towards higher values of $c/a$  and $b/a$ (see our Fig. A3 top panel and upper panel of Fig. 2 in \citealt{Zemp:2011ed}). This bias also depends on the weight chosen for the shape tensor. We chose no weight and \citet{Peter:2012jh} used a $r_{\rm ell}^{-2}$ weight. The latter weight results in a higher degree of sphericity compared to the former. According to our analysis (consistent with that of \citealt{Zemp:2011ed}), the difference is not very large. Without running a direct comparison using our code and the one used by \citet{Peter:2012jh} on the same set of haloes, we cannot pin down the exact reason for the difference. We emphasize, however, that in the most interesting regime for our analysis, beyond $10\%$ of the virial radius, our results agree with those of \citet{Peter:2012jh}.

The impact of baryonic physics in the halo shapes beyond the central regions of clusters could actually remain significant. According to results based on the Horizon-AGN simulation project, \citet{Suto:2017} conclude that the shapes of haloes can be affected at up to half of the virial radius, with an amplitude that is highly dependent on the bayonic physics added to the simulation, but that in general drives haloes to become more spherical, which mimics the effect of dark matter collisions. Nevertheless, in their current simulation with the full physics implementation (including strong AGN feedback), the impact on the distribution of halo shapes across a sample of simulated haloes is relatively mild at large radii. With a halo sample consisting of $40$ haloes with masses larger than $5\times10^{13}$~M$_\odot$, their results show a deviation between the distributions of the $c/a$ axis ratios of dark-matter-only and full-physics simulations is less than $\sim0.5\sigma$ at $R/R_{200}=0.1$ (see the right panel of Fig. 8 in \citealt{Suto:2017}), which is significantly less that the difference we measure between CDM and SIDM1 dark-matter-only simulations. Although to fully address this issue, it would be necessary to perform simulations with baryonic physics, these results are promising for the prospects of robustly differentiating the predictions for halo shapes from SIDM (at the level of 1~cm$^2$~gr$^{-1}$) and CDM at large radii based on dark-matter-only simulations.

\subsection{Velocity anisotropy}
\label{sec:beta}
Dark matter collisions not only affect the spatial distribution of dark matter haloes, but also their velocity distribution. In general, frequent collisions will lead to the thermalisation of the dark matter halo, making the velocity distribution Maxwellian and resulting in an isothermal velocity dispersion profile within the core of the SIDM haloes (e.g. \citealt{Vogelsberger:2012sa}). This process naturally results in isotropization of the orbits \citet{Hansen:2007jj}, which has not been analysed in most SIDM studies (see, however, Fig. 4 of \citealt{Vogelsberger:2014pda} and the discussion therein, as well as \citealt{Host:2008gi}). The degree of velocity anisotropy is given by the (radially dependent) parameter
\begin{align}
	\beta(r) = 1 - \frac{\sigma_{\phi}^2(r) + \sigma_{\theta}^2(r)}{2 \sigma_r^2(r)},
\end{align}
where $\sigma_r$, $\sigma_{\phi}$ and $\sigma_{\theta}$ are the velocity dispersions corresponding to the radial direction, and the azimuthal and zenith angles, respectively. As in the case of the density profile, we create spherical shells (equidistant in logarithmic scale) centred at the minimum of the potential of a given halo, and calculate the average velocity dispersions in each direction. We have removed substructures to concentrate on the velocity anisotropy of the main (smooth) halo component, but this choice does not change our results significantly.

\begin{figure}
	\centering
	\includegraphics[width=\linewidth]{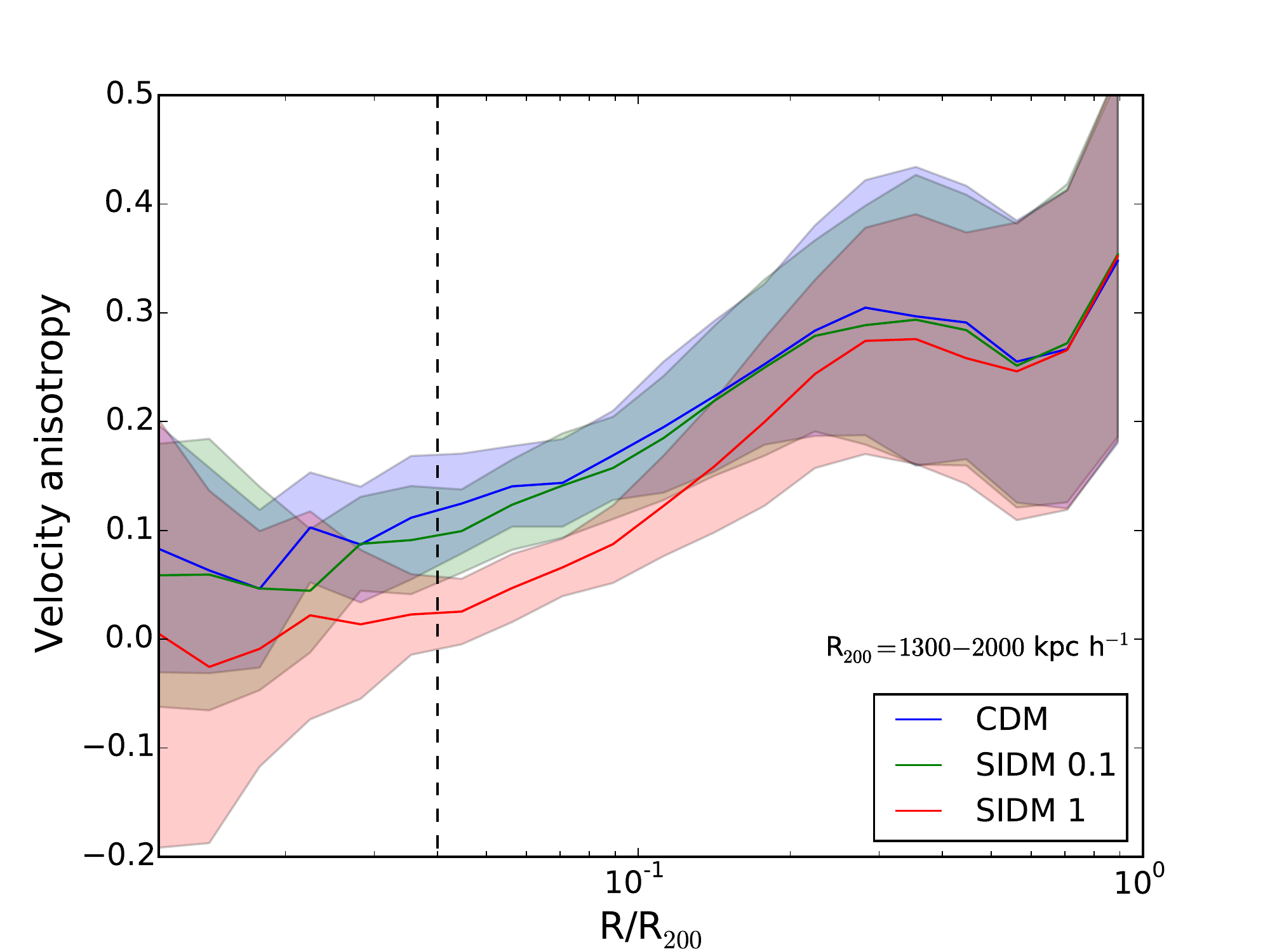}
	\caption{The combined velocity dispersion profiles for the three cosmologies CDM (blue), SIDM0.1 (green) and SIDM1 (red). The solid lines indicate the medians of the distributions, whereas the shaded regions show the 1$\sigma$ confidence intervals. The black dashed line marks the radius above which we trust the results. We see a small difference in the median and scatter of the distributions between CDM and SIDM0.1 near the trust radius of 0.04~R$_{200}$, and a significant separation (at the $1\sigma$ level) between CDM and SIDM1 from the trust radius to about 0.1~R$_{200}$.}
	\label{fig:results:betaprofiles1}
\end{figure}

\begin{figure}
	\centering
	\includegraphics[width=\linewidth]{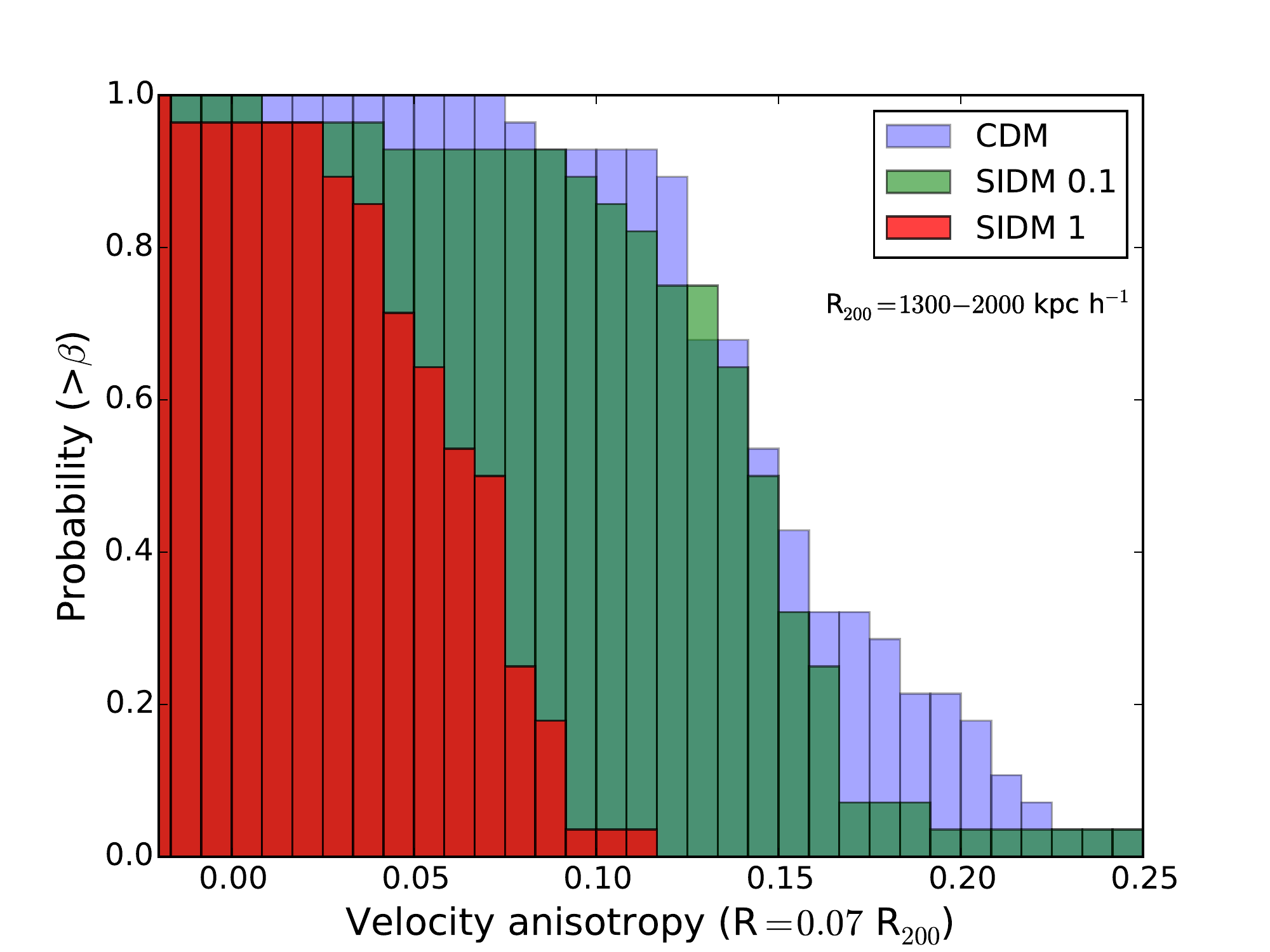}
	\caption{Cumulative distribution of the velocity anisotropy $\beta$ at 0.07~R$_{200}$ ($\sim100-140$~kpc~h$^{-1}$), showing the probability that $\beta$ is larger than a given value for CDM (blue), SIDM0.1 (green) and SIDM1 (red). We see that for CDM, all but two of the haloes have $\beta \gtrsim 0.1$, whereas only one SIDM1 halo has a velocity anisotropy above this value. Likewise we see significantly more haloes in the high velocity anisotropy tail of the CDM distribution ($\sim25$\% or $7-8$ haloes at about $\beta=0.175$) compared to SIDM0.1 ($<10$\% or 2 haloes).}
	\label{fig:results:betahist}
\end{figure}

Fig.~\ref{fig:results:betaprofiles1} shows the radial velocity anisotropy profiles for our combined halo sample using the same color code as in previous figures. It is well established that $\beta$ is close to zero in the inner regions of CDM haloes (see e.g. \citealt{Lemze:2011ud}, \citealt{Sparre:2012} and \citealt*{Wojtak:2013eia}). In the SIDM case, $\beta$ also tends to zero in the centre but more rapidly than CDM and through a different process, that of collisional relaxation. Very close to the halo centre, the increase in numerical noise clearly becomes a relevant issue and we cannot trust our results below $0.04$~R$_{200}$. Beyond this trust radius, we find that the median of the distribution of $\beta$ values is consistently smaller in SIDM compared to CDM at \textit{all but the very largest radii.} The distribution itself is narrower in the former than in the latter.

Particularly, we find that in the intermediate region around $(4-10)\%$ of the virial radius (corresponding to $\sim 60-150$~kpc~h$^{-1}$) there is a clear difference in the median and scatter of the velocity anisotropy distribution between CDM and SIDM haloes. Any intrinsic scatter in velocity anisotropy across haloes due to divergent assembly history is reduced considerably once dark matter collisions are present. A particular consequence of this is that in a SIDM0.1 cosmology, we find small likelihood ($<10\%$, compared to $\sim25\%$ for CDM) of finding massive cluster-sized haloes with $\beta \gtrsim 0.175$ at 0.07~R/R$_{200}$ (see Figure \ref{fig:results:betahist}). The difference between CDM and SIDM becomes more pronounced with increased cross section, and we would find almost no haloes with a $\beta \gtrsim 0.1$ at 0.07~R/R$_{200}$ for SIDM1, while almost all CDM haloes have $\beta\gtrsim0.1$ at this radius. Looking for such systems offers an interesting alternative way to constrain SIDM models. Consider for instance that we observe 25 CDM haloes (of the 28 in total) with $\beta>0.12$ at this radius (see Figure \ref{fig:results:betahist}). For such a number count, the Poisson lower and upper limits at the corresponding Gaussian $3\sigma$ level are 12.6 and 44, respectively \citep{Gehrels:1986}. Since there are no haloes in the SIDM1 case with $\beta>0.12$, this is a promising avenue that could lead to competitive constraints on SIDM. This estimation is of course given without considering the potentially substantial errors in the measurement of $\beta$. The biggest uncertainty originates from the cluster to cluster dispersion of the calibration of the dark matter to gas temperature ratio, which is done against numerical simulations. Currently this ratio has a dispersion of approximately $10\%$. When this is folded with the observational uncertainty on the gas temperature, which is typically about $10\%$ for well-observed clusters, this leads to error-bars on the dark matter velocity anisotropy with an amplitude between 0.05 and 0.15, with the smallest error-bars being within $r_{500}$ (where the average density is 500 times the critical density of the universe) (Svensmark et al., in prep). The largest separation between the medians of the distribution in $\beta$ in CDM and SIDM1 is $\sim0.1$, which means that current observational uncertainties are still too large to use the velocity anisotropy as a diagnostic to constrain SIDM. It is expected that near future numerical simulations will allow the calibration of the dispersion on the temperature ratios to be improved by approximately a factor of two, while the X-ray observations of massive clusters using {\it XMM-Newton} and {\it Chandra} continuously decrease the uncertainties on measured gas temperatures.

\begin{figure}
	\centering
	\includegraphics[width=\linewidth]{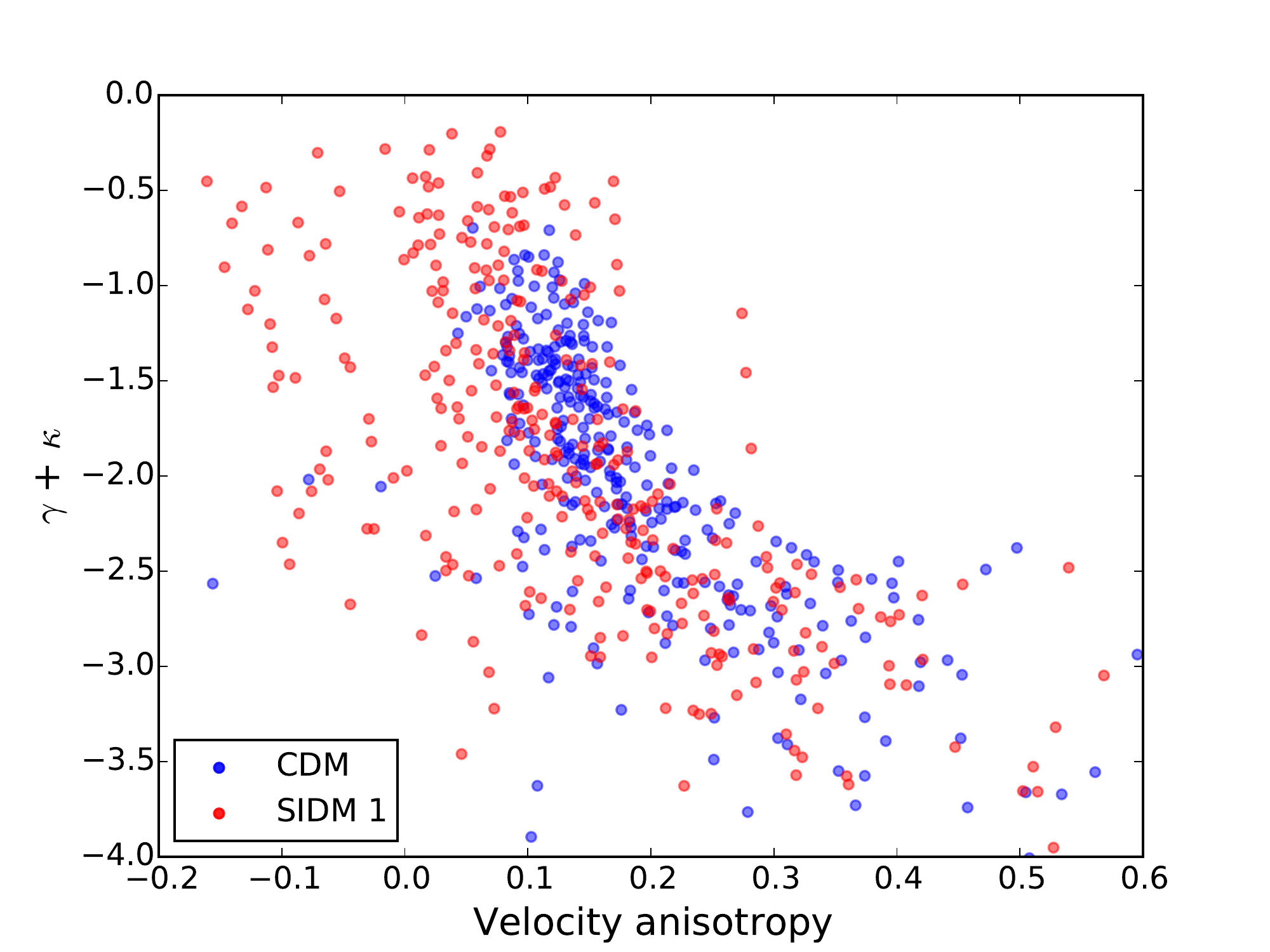}
	\caption{Sum of the logarithmic derivatives of the density and squared radial velocity dispersion profiles ($\gamma + \kappa$) vs velocity anisotropy. The data shows all 28 haloes in 11 different radial bins for CDM (blue) and SIDM1 (red).}
	\label{fig:results:gammakappa}
\end{figure}

\citet{Hansen:2004qs} showed that there is a correlation between the logarithmic slope of the density profile $\gamma(r) = d {\rm log_{10}} \rho / d {\rm log_{10}} r$ and the velocity anisotropy profile $\beta(r)$ for relaxed structures in a CDM cosmology, which has a non-linear behaviour (see e.g. Fig. 11 \citealt{Sparre:2012}). In the case of SIDM we find that the relation becomes tighter and closer to linear due to the creation of the central core. 
\citet*{Hansen:2010kb} showed that by relating the sum of the logarithmic derivatives of the density and squared radial velocity dispersion profiles ($\gamma + \kappa$) to the velocity anisotropy, where $\kappa = d {\rm log_{10}}(\sigma_{r}^{2}) / d {\rm log_{10}} r$, it is possible to simplify the Jeans equation to
\begin{align}\label{eq_s}
\frac{G M_{\rm tot}}{r} &= - \sigma_{r}^{2} (\gamma + \kappa + 2 \beta)= - \sigma_{r}^{2} \rm A (\gamma + \kappa).
\end{align}
For SIDM1 we find that $(\gamma + \kappa) \sim -24\beta$ for the two innermost radial bins outside the trust radius, corresponding to $A\sim11/12$ in eq.~(\ref{eq_s}). The connection between $(\gamma + \kappa)$ and $\beta$ can be seen in Figure \ref{fig:results:gammakappa} individually for each of our 28 haloes in 11 different radial bins (all above the trust radius in our simulation). The figure shows differences between CDM and SIDM1, particularly in the inner region (top left). Following the approach of \citet{Hansen:2010kb}, we can attempt to quantify this difference by comparing an effective equation of state for each model, $P = b$~$(\rho \sigma_{r}^{2})^{K}$, where $b$ is a constant and $K$ is the polytropic index. If we require the Jeans equation to take the form of the hydrostatic equilibrium equation, we can then interpret the behavior of the dark matter as that of a fluid with a given polytropic index. For CDM,  \citet{Hansen:2010kb} found an effective polytropic index $K_{\rm CDM}=3/4$.\footnote{Compare this to e.g. 1 for an isothermal gas or 5/3 for an ideal monatomic gas.} We find a similar result with $K_{\rm CDM}=4/5$, and find that for SIDM1 the effective polytropic index for the inner region is $K_{\rm SIDM}\sim11/12$. As expected for a self-interacting dark matter species, this is an index closer to one than for pure collisionless dark matter, while still not in the collisional fluid regime.
\subsection{The memory of assembly history in SIDM haloes}
\label{sec:assembly}
Although the general expectation is that dark matter collisions will tend to erase the memory of the assembly process in the inner regions of haloes, the extent to which this process is complete depends on the amplitude of the scattering cross section. For instance, for 1~cm$^2$~gr$^{-1}$, \citet{Peter:2012jh} noted that at a fixed $V_{\rm max}$ (a proxy for halo mass) a scatter of a factor of $\sim3$ in central densities is observed in SIDM simulations, suggesting that a signature of the assembly history remains in the distribution of central densities in SIDM haloes with cross sections of this order of magnitude. This signature has not been studied in detail and in this section we present such an analysis based on our cluster-sized simulations. 

To get a simple measure of the mass assembly history of our halo sample, we trace back in time each halo from $z=0$ through successive snapshots, picking the most likely progenitor based on proximity of mass and distance from the selected halo in consecutive snapshots. From this excercise, we define a measure of the ``formation'' time of a halo as the redshift $z_{f}$ where it acquires a fraction $f_{\rm mass}$ of the mass contained within the scale radius $r_{-2}$ at $z=0$, where $r_{-2}$ is the radius at which the logarithmic slope of the density profile is $-2$:\footnote{We note that using a fraction of $M_{200}$ instead of $M(r_{-2})$ to define the formation redshift leads to qualitatively similar results.}
\begin{align}\label{eq_form}
M_{200}(z_f) = M(r < f_{\rm mass} \times r_{-2}, z=0)
\end{align}
This definition gives us a way to explore if the dispersion in the central densities at a fixed $V_{\rm max}$ is related to the time when the mass in the inner regions of haloes was assembled. We note that using eq.~(\ref{eq_form}) for small values of $f_{\rm mass}$ leads to a biased definition of formation time in the SIDM case since once within the region of the SIDM core, the enclosed mass is clearly smaller in the SIDM case than in the CDM case, leading to an artificially larger formation time in the former. In fact, SIDM and CDM haloes assemble at roughly the same time, since self-interactions start to be important later in the evolution of haloes. Eq.~(\ref{eq_form}) is nevertheless sufficient and simple enough for the purpose of this section. We have verified that using the formation time assigned to CDM haloes to the corresponding SIDM haloes leaves our results qualitatively unchanged.

\begin{figure}
\centering
\includegraphics[width=\linewidth]{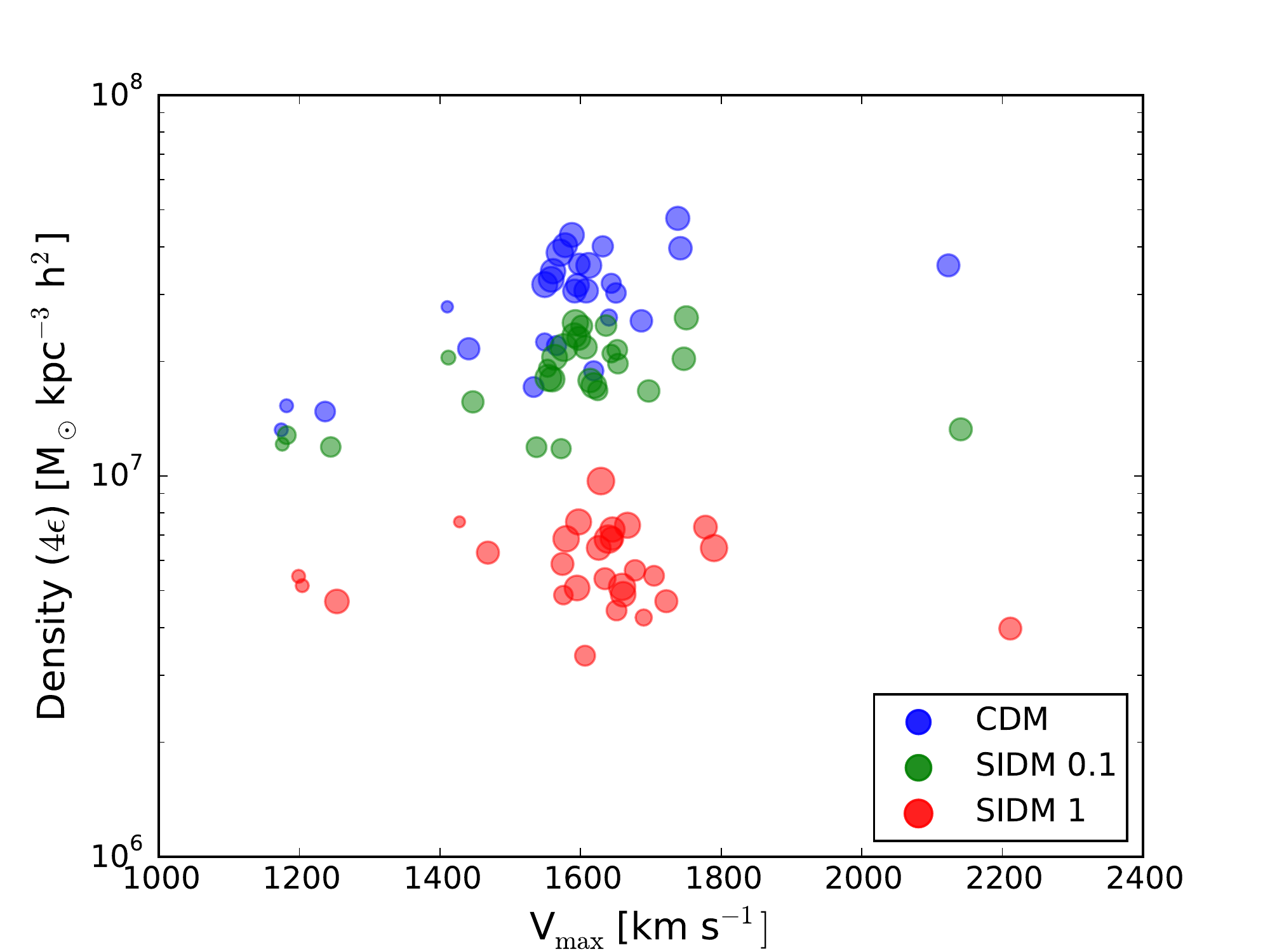}
\caption{Density at four times the softening ($\approx 22$~kpc~h$^{-1}$) vs the maximum velocity of the halo. The type of simulation is shown by the color (blue for CDM, green for SIDM0.1 and red for SIDM1) and the symbol size indicates the square of the formation redshift (where the halo had obtained half its mass within $M(r_{-2})$), i.e., smaller (larger) symbols indicate haloes formed later (earlier). For CDM and SIDM0.1 we see a mild trend between the central density and the maximum velocity of a halo, which is absent negligible in the case of SIDM1.}
\label{fig:results:assembly2}
\end{figure}

Fig.~\ref{fig:results:assembly2} shows the density at a fixed radius (4$\epsilon$) as a function of V$_{\rm max}$ (a proxy of halo mass) for our halo sample in the different cosmologies. We have chosen $r=4\epsilon \approx 22$~kpc~h$^{-1}$ as the minimum radius we can safely trust in our density profiles. The size of each symbol is proportional to the formation redshift $z_{f}$ (defined with $f_{\rm mass}=1/2$), with larger symbols corresponding to haloes formed earlier. It is well known that in CDM, at a fixed radius, more massive haloes (larger V$_{max}$) are denser than less massive ones, while at a fixed mass (V$_{max}$), more concentrated haloes form earlier than less concentrated ones. Both of these features are present in Fig.~\ref{fig:results:assembly2}, although the trends are noisier due to the limited mass range we cover and the low number of haloes in our sample. 

The trend between central density and $V_{\rm max}$ is weaker in the SIDM 0.1 case, and it essentially disappears in SIDM 1, where there are even hints of an inversion in the relation, albeit the limitations of our sample size prevent a clear conclusion in this regard. This can be quantified by Spearman's rank correlation coefficient, for which we get 0.538, 0.386, -0.097 (CDM, SIDM 0.1, SIDM1). We notice that this trend is consistent with the one found in \citet{Rocha:2012jg}, and explained on the basis of the toy model developed there in which the relevant quantity is a characteristic radius (where the average scattering per particle over the history of the halo is one) that separates the inner region, where scattering is effective, from the outer region, which follows CDM (see Eq.~19 and Fig.~12 of \citealt{Rocha:2012jg}). 

\begin{figure}
	\centering
	\includegraphics[width=\linewidth]{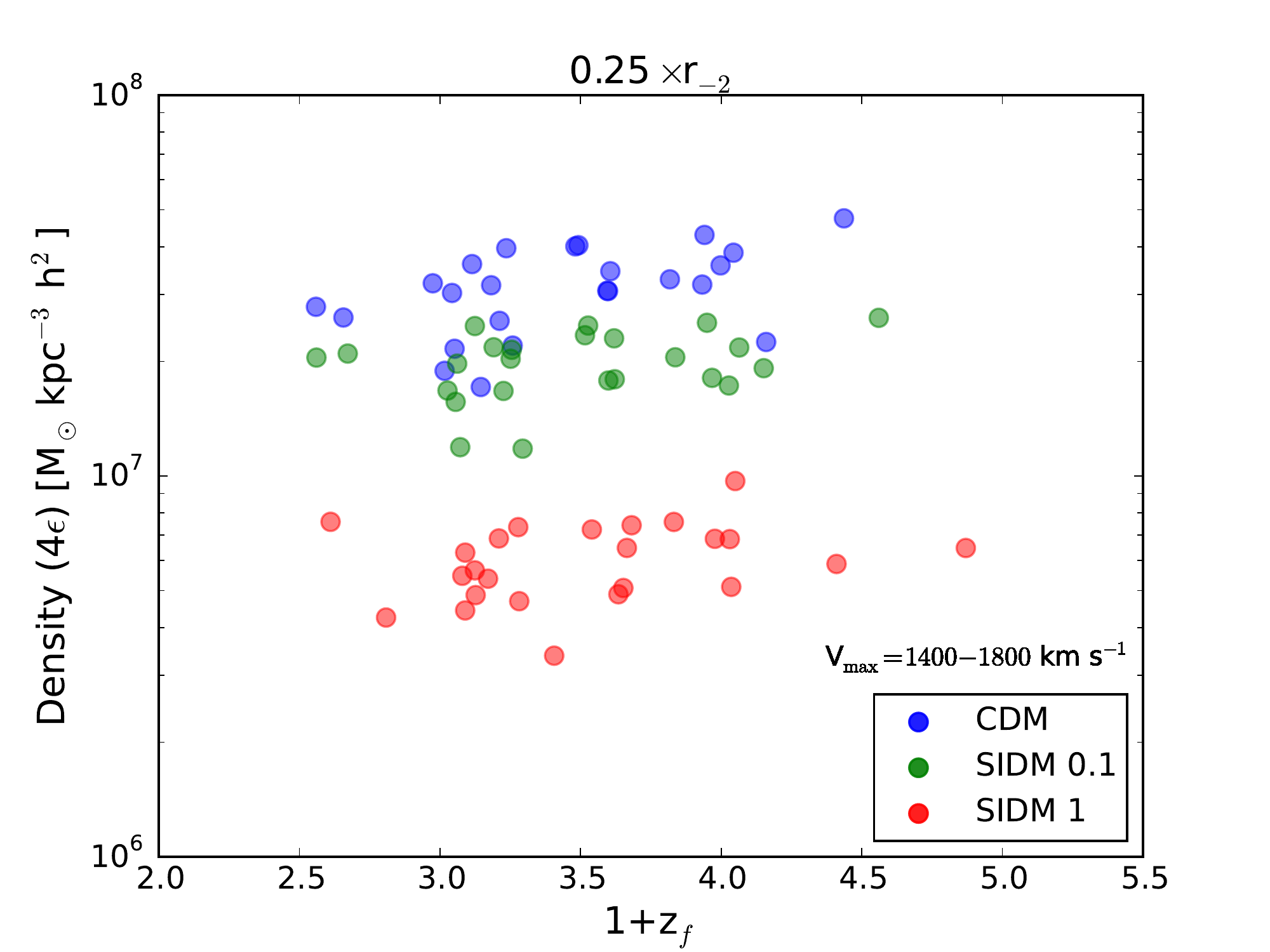}
	\caption{The density at four times the softening ($\approx 22$~kpc~h$^{-1}$) vs the formation redshift, given by the redshift where the halo has obtained a virial mass, M$_{200}(z_f)$, equal to the mass within $0.25 \times$r$_{-2,\rm CDM}$ at $z=0$. Only a subsample of haloes with V$_{\rm max}$ between 1400 and 1800 km/s are shown.}
	\label{fig:results:assembly3}
\end{figure}

The weak trend between density at $4\epsilon$ and formation redshift can be appreciated in Fig.~\ref{fig:results:assembly3}, where we have restricted the halo sample to $1400~{\rm km/s}\leq V_{\rm max}\leq1800$~km/s and we define the formation redshift according to a more internal region of the halo, taking $f_{\rm mass}=1/4$ instead of a half. In this plot, all cases show a very similar weak trend: for haloes of similar mass, the earlier they form, the denser they are.

\begin{figure}
	\centering
	\includegraphics[width=\linewidth]{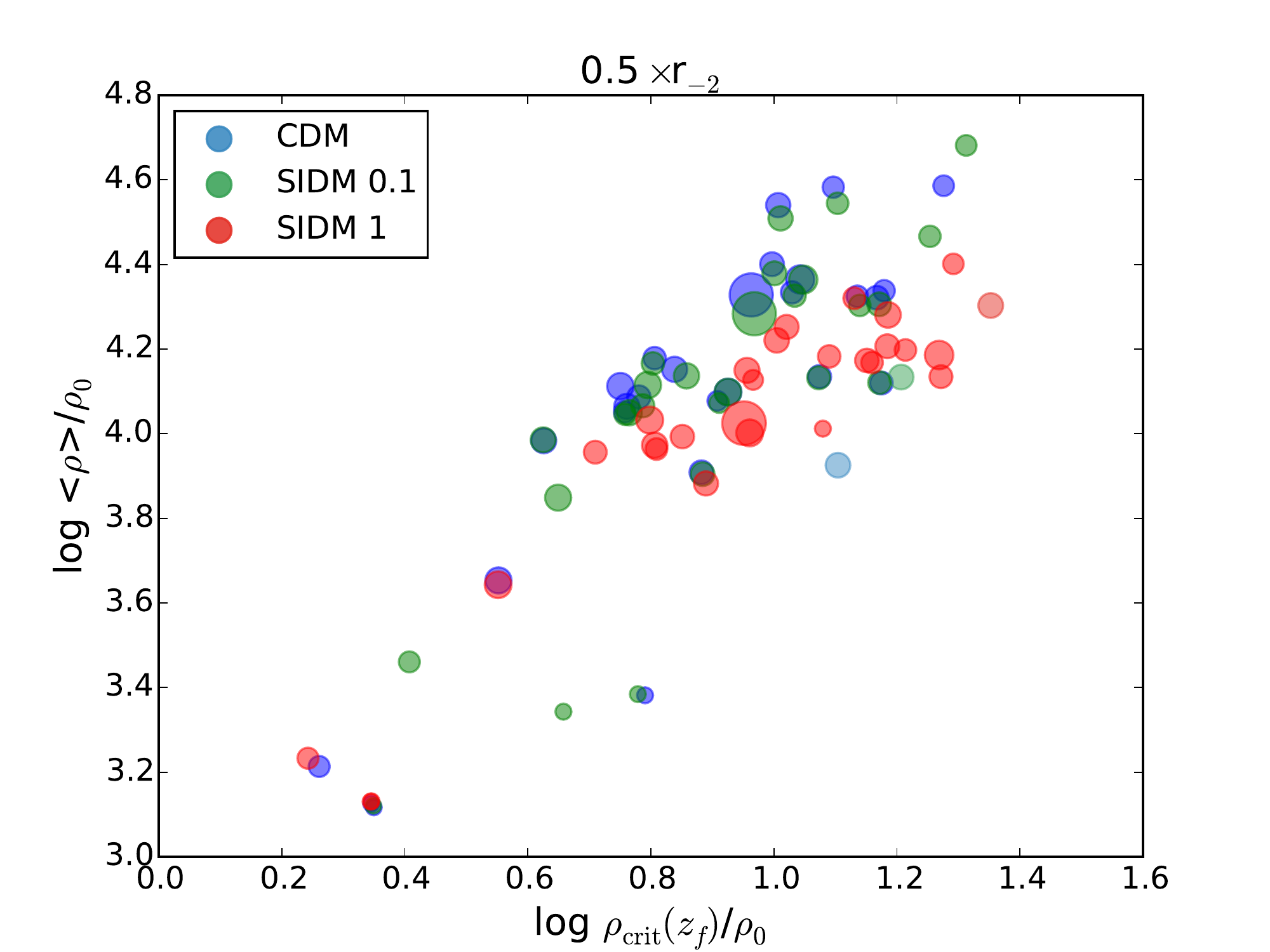}\\
	\includegraphics[width=\linewidth]{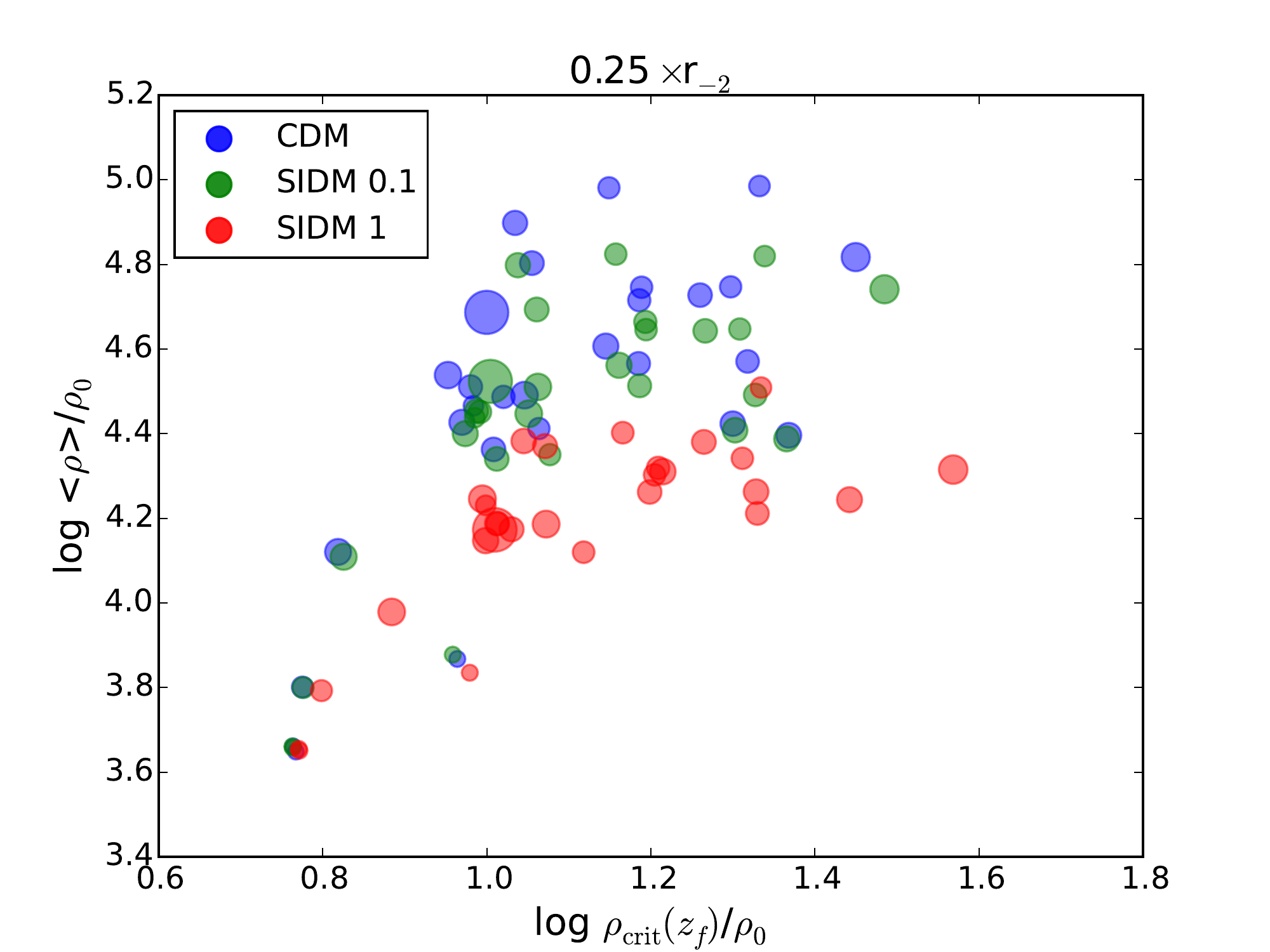}
	\caption{\textbf{Top:} Mean density within 0.5$\times$r$_{-2}$ scaled by the critical density today vs the critical density at formation redshift (the redshift where the halo has obtained a virial mass equal to the mass enclosed within 0.5$\times$r$_{-2}$ at $z=0$), likewise scaled by the critical density today. Larger symbol size indicates greater virial mass M$_{200}$ at $z=0$, scaled to the power 1.5. \textbf{Bottom:} Similar to above, except that the mean density is within a smaller radius of 0.25$\times$r$_{-2,\rm CDM}$, this time given by the coresponding CDM halo in each case.}
	\label{fig:results:assembly4}
\end{figure}

To carry out a deeper exploration of the relation between the assembly history and the scatter of the central halo densities, we follow the methodology of \citet{Ludlow:2013bd} that connects the mass assembly history with the mass profile of dark matter haloes. In particular, we relate the mean inner halo density (within $r_{-2}/2$ and $r_{-2}/4$) with the critical density of the Universe at the redshifts when the virial mass of the main progenitor was equal to $M(r_{-2}/2)$ and $M(r_{-2}/4)$, which are the formation redshifts we explored above. Fig.~\ref{fig:results:assembly4} shows the connection between these two quantities in our halo sample for the different cosmologies (same color code as in previous figures). For CDM, we confirm the results of \citet{Ludlow:2013bd} where these two quantities are intimately related, with haloes that assembled earlier being the most centrally dense. 

For the population as a whole, we see that there is a deviation of the SIDM1 behaviour relative to the CDM case: the higher the value of $\rho(z_f)$, i.e., the earlier the halo formed, the lower the value for $r_{-2}$ (i.e. the higher the concentration of the CDM halo), and therefore there is more time for self-interactions to have an impact, resulting in a stronger reduction in the inner density $<\rho(r_{-2}/2)>$. This effect manifests as a flattening in the right side of the correlation in the top panel of Fig.~\ref{fig:results:assembly4}. In other words, the region that $<\rho(r_{-2}/2)>$ is probing is closer to the center as the formation redshift gets higher. This is why CDM and SIDM are quite close to each other on the lower left, and are strongly separated on the upper right. This effect is more dramatic in the bottom panel of Fig.~\ref{fig:results:assembly4}, where the formation time is defined from a more central region ($f_{\rm mass}=1/4$).

Notice also that, in this case, the formation redshifts for SIDM1 are clearly higher than in CDM, since the values of $M(r_{-2}/4)$ are considerably lower in the former than in the latter and therefore artificially associated to a more ancient formation time (as discussed earlier in this section). If we instead consider the formation redshift as defined by $M_{200}(z_f) =$~$ M_{\rm CDM}(r <$~$ f_{\rm mass}$~$\times$~$ r_{-2}, z =$~$ 0)$, we unsurprisingly find only a small shift in formation time between CDM and SIDM. This however, does not change the overall conclusion drawn from Fig.~\ref{fig:results:assembly4}.

We also note that the formation history has an impact on the shape of our haloes, in that haloes having formed earlier are more spherical. This effect is more pronounced in SIDM compared to CDM, with a $\sim 10-15$\% increase in sphericity for SIDM1 haloes in the formation redshift range explored (z$_f \sim 1-2$), which makes sense considering there is more time for collisions to isotropize the orbits.

\section{Summary and conclusions}
\label{sec:conclusions}

Among the possible dark matter interactions that can substantially change structure formation in the Universe, strong self-scattering (SIDM) remains one of the least constrained. The consequences of SIDM only become evident indirectly, through its dynamical impact in the central region of haloes and the galaxies within. However, these regions are also the places where the dynamical impact of gas cooling, star formation, supernova, and AGN feedback (baryonic physics) is also substantial. The predicted final dark matter distribution within the scale of galaxies can become degenerate between strong dark matter self-interactions and strong feedback (from AGN or SNe), with both predicting a substantial reduction of the inner halo densities over the naive predictions of a pure Cold Dark Matter model (without baryonic physics). This situation makes the prospect of identifying the dynamical signature of dark matter collisions within the centers of galaxies a challenging task. 

Nonetheless, since there is only a narrow window for a {\it constant} cross section SIDM model to be a compelling alternative to the Cold Dark Matter model for structure formation, ($0.1$~cm$^2$~gr$^{-1}<\sigma/m_\chi\lesssim$ 1~cm$^2$~gr$^{-1}$), there is a case to search for definitive predictions that can separate CDM from SIDM. In this paper, we aim at contributing to this effort by performing dark-matter-only simulations to analyse in detail the structure of haloes in both cosmologies at large radii, far from the strongest impact of baryonic physics. Although the effects of SIDM are also diminished at large radii, our results quantify important differences between CDM and SIDM even in this regime.

Our analysis is based on a sample of 28  {\it relaxed} massive cluster-sized haloes ($\sim10^{15}$~M$_\odot$) simulated under the same initial conditions for CDM and SIDM with two cross sections, $\sigma/m_\chi=0.1$~cm$^2$~gr$^{-1}$ (SIDM~0.1) and $1$~cm$^2$~gr$^{-1}$ (SIDM~1). These are zoom simulations (from a larger 1~Gpc$^3$ parent simulation) consistent with a Planck cosmology with a softening length $\epsilon\sim8$~kpc, and particle mass $\sim10^9$~M$_\odot$. Our main findings are summarised in the following (unless otherwise stated, these results are for $z=0$).
\begin{itemize}
\item At the level of $\sigma/m_\chi=1$~cm$^2$~gr$^{-1}$, SIDM has a global impact on the virial ratio ($2T/\vert U\vert$) of a halo. We assessed this using our larger parent simulation, finding that for $z<1$ the median of the distribution of virial ratios is lower in SIDM~1 than in CDM by $0.5-1\%$. Albeit small, this difference implies that a larger fraction of haloes are predicted to be {\it relaxed} in SIDM~1, relative to CDM (see Table \ref{tab:relaxation} and Section \ref{sec:relaxation}). 
\item The distribution of density profiles differs between CDM and SIDM~1 at the $1\sigma$ level already at 5\% of the virial radius ($\sim110-150$~kpc, depending on the mass of the cluster). The logarithmic slope of the density profiles shows an even larger difference extending towards larger radii, with the distributions separating at the $1\sigma$ level at 10\% of the virial radius (see Figs.~\ref{fig:results:density}-\ref{fig:results:slope}). At these radii, the impact from baryonic physics is expected to be quite below the differences between the dark matter models \citep[e.g.][]{Schaller:2015,Peirani:2016}.
\item Comparing our density profiles to the observationally inferred density profiles of a sample of 7 galaxy clusters from \citet{Newman:2012nw}, we see a strong indication that SIDM~1 is disfavoured by observations at an intermediate radial range: $2-5\%$ of the virial radius (see Fig. \ref{fig:results:density_newman}). However, there is an important caveat. The impact of baryons on a SIDM halo at these scales is non-trivial and must be studied carefully with simulations to reach a more definitive conclusion.
\item The distribution of halo shape profiles, characterised by the radial profiles of the axis ratios $c/a$ and $b/a$ of ellipsoidal shapes, differs between CDM and SIDM~1 at the $1\sigma$ level across a large radial range between 5\% and 30\% of the virial radius ($\sim150-740$~kpc, depending on the mass of the cluster). The impact of baryonic physics in halo shapes is rather uncertain, but in a recent state-of-the-art simulation with a strong AGN feedback, the impact of baryonic physics at 10\% of the virial radius is $\sim0.5\sigma$ relative to the CDM case without baryonic physics \citep{Suto:2017}, quite below the differences we are reporting between SIDM~1 and CDM. 
\item The distribution of velocity anisotropy profiles differs between CDM and SIDM at the $1\sigma$ level below $10\%$ of the virial radius, with the SIDM distribution skewed towards values closer to $\beta=0$ (see Fig.~\ref{fig:results:betaprofiles1}). At relatively large radii (7\% of the virial radius, $140-200$kpc, depending on halo mass) we found no SIDM~1 haloes with $\beta\gtrsim0.12$, compared to nearly $90\%$ of CDM haloes (see Fig.~\ref{fig:results:betahist}).
Given the expectation of a small impact of baryonic physics at these radii, we estimate that an observational sample of $\sim30$ relaxed clusters, in a similar mass range as the one explored here, could put competitive constraints on SIDM using the values of $\beta$ at large radii as a diagnostic.
\item For all of these diagnostics, at the level of $\sigma/m_\chi=0.1$~cm$^2$~gr$^{-1}$ the differences between CDM and SIDM are too small to draw significant conclusions. Separating the models in this limiting cross section requires getting quite close to the inner region of the haloes, and thus it entails taking into account the full impact of baryonic physics.
\end{itemize}

We conclude that significant structural differences remain between SIDM and CDM at the level of a cross section per unit mass of 1~cm$^2$~gr$^{-1}$, even at very large radii $\gtrsim10\%$ of the virial radius, where the impact of baryonic physics is expected to be limited and less than that of these two different dark matter cases. Because of this, our results (based on dark-matter-only simulations) offer guidelines for observational comparisons with relaxed massive galaxy clusters.  

In addition, to explore the structural differences between SIDM and CDM at large radii, we also studied whether a memory of the assembly history remains in SIDM haloes, as it does in the CDM case. This has not been analysed before in detail, with the paper by \citet{Peter:2012jh} being the most relevant reference. Remarkably we find that, at the level of 1~cm$^2$~gr$^{-1}$, the memory of assembly history in massive cluster sized haloes remains in SIDM to some extent. In particular, by exploring the central densities of our simulated haloes, we find that (see Figs.~\ref{fig:results:assembly2}$-$\ref{fig:results:assembly3}): (i) at a fixed radius, more massive haloes are denser in CDM, while in SIDM~1 this trend disappears; (ii) at a fixed mass, denser haloes formed earlier in CDM, with a similar trend for SIDM~1. To show the differences between CDM and SIDM~1 in a clear way, we used the connection between the mass assembly history and halo mass profiles in CDM reported by \citet{Ludlow:2013bd}. This connection is reflected in a tight correlation between the average inner halo density (within a fraction $f$ of $r_{-2}$, the scale radius for the NFW profile) at $z=0$, and the density of the Universe at the time when the mass of the main progenitor of the halo was equal to $M(fr_{-2})$. In the case of SIDM~1, this correlation is broken in the central regions within the density core (see Fig.~\ref{fig:results:assembly4}), since for haloes formed earlier there has been more time for dark matter collisions to thermalize the central region and reach a stable (previous to the stage of gravothermal collapse) constant central value of the density. Haloes above this threshold of thermalization have thus very similar central densities, regardless of their formation time.  The memory of assembly history in SIDM is thus indeed erased only within the very central regions of the halo ($r_{-2}/4$), and only for those haloes that have crossed this formation redshift threshold $z_f>2$. 

\section*{Acknowledgements}
We thank Anastasia Sokolenko, Kyrylo Bondarenko and Alexey Boyarsky for useful discussions related to the halo velocity dispersion profiles that led to an improvement of our paper. We also thank Hai-Bo Yu for useful discussions and suggestions to improve our work. JZ acknowledges support by a Grant of Excellence from the Icelandic Research Fund (grant number 173929$-$051). DR is supported by a NASA Postdoctoral Program Senior Fellowship at the NASA Ames Research Center, administered by the Universities Space Research Association under contract with NASA. SHH is partially funded by the Danish council for independent research, DFF 6108-00470. MV acknowledges support through an MIT RSC award, the support of the Alfred P. Sloan Foundation, and support by NASA ATP grant NNX17AG29G.

\bibliography{references}

\begin{thebibliography}{}
\makeatletter
\relax
\def\mn@urlcharsother{\let\do\@makeother \do\$\do\&\do\#\do\^\do\_\do\%\do\~}
\def\mn@doi{\begingroup\mn@urlcharsother \@ifnextchar [ {\mn@doi@}
  {\mn@doi@[]}}
\def\mn@doi@[#1]#2{\def\@tempa{#1}\ifx\@tempa\@empty \href
  {http://dx.doi.org/#2} {doi:#2}\else \href {http://dx.doi.org/#2} {#1}\fi
  \endgroup}
\def\mn@eprint#1#2{\mn@eprint@#1:#2::\@nil}
\def\mn@eprint@arXiv#1{\href {http://arxiv.org/abs/#1} {{\tt arXiv:#1}}}
\def\mn@eprint@dblp#1{\href {http://dblp.uni-trier.de/rec/bibtex/#1.xml}
  {dblp:#1}}
\def\mn@eprint@#1:#2:#3:#4\@nil{\def\@tempa {#1}\def\@tempb {#2}\def\@tempc
  {#3}\ifx \@tempc \@empty \let \@tempc \@tempb \let \@tempb \@tempa \fi \ifx
  \@tempb \@empty \def\@tempb {arXiv}\fi \@ifundefined
  {mn@eprint@\@tempb}{\@tempb:\@tempc}{\expandafter \expandafter \csname
  mn@eprint@\@tempb\endcsname \expandafter{\@tempc}}}

\bibitem[\protect\citeauthoryear{Ackerman, Buckley, Carroll  \&
  Kamionkowski}{Ackerman et~al.}{2009}]{Ackerman:mha}
Ackerman L.,  Buckley M.~R.,  Carroll S.~M.,   Kamionkowski M.,  2009, \mn@doi
  [Phys. Rev.] {10.1103/PhysRevD.79.023519, 10.1142/9789814293792_0021}, D79,
  023519

\bibitem[\protect\citeauthoryear{Arkani-Hamed, Finkbeiner, Slatyer  \&
  Weiner}{Arkani-Hamed et~al.}{2009}]{ArkaniHamed:2008qn}
Arkani-Hamed N.,  Finkbeiner D.~P.,  Slatyer T.~R.,   Weiner N.,  2009, \mn@doi
  [Phys. Rev.] {10.1103/PhysRevD.79.015014}, D79, 015014

\bibitem[\protect\citeauthoryear{{Benson}}{{Benson}}{2017}]{2017MNRAS.471.2871B}
{Benson} A.~J.,  2017, \mn@doi [\mnras] {10.1093/mnras/stx1804}, \href
  {http://adsabs.harvard.edu/abs/2017MNRAS.471.2871B} {471, 2871}

\bibitem[\protect\citeauthoryear{Bett, Eke, Frenk, Jenkins, Helly  \&
  Navarro}{Bett et~al.}{2007}]{Bett:2006zy}
Bett P.,  Eke V.,  Frenk C.~S.,  Jenkins A.,  Helly J.,   Navarro J.,  2007,
  \mn@doi [Mon. Not. Roy. Astron. Soc.] {10.1111/j.1365-2966.2007.11432.x},
  376, 215

\bibitem[\protect\citeauthoryear{Buckley \& Fox}{Buckley \&
  Fox}{2010}]{Buckley:2009in}
Buckley M.~R.,  Fox P.~J.,  2010, \mn@doi [Phys. Rev.]
  {10.1103/PhysRevD.81.083522}, D81, 083522

\bibitem[\protect\citeauthoryear{Colin, Avila-Reese, Valenzuela  \&
  Firmani}{Colin et~al.}{2002}]{Colin:2002nk}
Colin P.,  Avila-Reese V.,  Valenzuela O.,   Firmani C.,  2002, \mn@doi
  [Astrophys. J.] {10.1086/344259}, 581, 777

\bibitem[\protect\citeauthoryear{{Cyr-Racine}, {Sigurdson}, {Zavala},
  {Bringmann}, {Vogelsberger}  \& {Pfrommer}}{{Cyr-Racine}
  et~al.}{2016}]{Cyr-Racine:2016}
{Cyr-Racine} F.-Y.,  {Sigurdson} K.,  {Zavala} J.,  {Bringmann} T.,
  {Vogelsberger} M.,   {Pfrommer} C.,  2016, \mn@doi [\prd]
  {10.1103/PhysRevD.93.123527}, \href
  {http://adsabs.harvard.edu/abs/2016PhRvD..93l3527C} {93, 123527}

\bibitem[\protect\citeauthoryear{{Dav{\'e}}, {Spergel}, {Steinhardt}  \&
  {Wandelt}}{{Dav{\'e}} et~al.}{2001}]{Dave:2001}
{Dav{\'e}} R.,  {Spergel} D.~N.,  {Steinhardt} P.~J.,   {Wandelt} B.~D.,  2001,
  \mn@doi [\apj] {10.1086/318417}, \href
  {http://adsabs.harvard.edu/abs/2001ApJ...547..574D} {547, 574}

\bibitem[\protect\citeauthoryear{{Dubois} et~al.,}{{Dubois}
  et~al.}{2014}]{Dubois:2014}
{Dubois} Y.,  et~al., 2014, \mn@doi [\mnras] {10.1093/mnras/stu1227}, \href
  {http://adsabs.harvard.edu/abs/2014MNRAS.444.1453D} {444, 1453}

\bibitem[\protect\citeauthoryear{{Elbert}, {Bullock}, {Garrison-Kimmel},
  {Rocha}, {O{\~n}orbe}  \& {Peter}}{{Elbert} et~al.}{2015}]{Elbert2015}
{Elbert} O.~D.,  {Bullock} J.~S.,  {Garrison-Kimmel} S.,  {Rocha} M.,
  {O{\~n}orbe} J.,   {Peter} A.~H.~G.,  2015, \mn@doi [\mnras]
  {10.1093/mnras/stv1470}, \href
  {http://adsabs.harvard.edu/abs/2015MNRAS.453...29E} {453, 29}

\bibitem[\protect\citeauthoryear{{Elbert}, {Bullock}, {Kaplinghat},
  {Garrison-Kimmel}, {Graus}  \& {Rocha}}{{Elbert} et~al.}{2016}]{Elbert:2016}
{Elbert} O.~D.,  {Bullock} J.~S.,  {Kaplinghat} M.,  {Garrison-Kimmel} S.,
  {Graus} A.~S.,   {Rocha} M.,  2016, preprint, \href
  {http://adsabs.harvard.edu/abs/2016arXiv160908626E} {} (\mn@eprint {arXiv}
  {1609.08626})

\bibitem[\protect\citeauthoryear{Feng, Kaplinghat, Tu  \& Yu}{Feng
  et~al.}{2009}]{Feng:2009mn}
Feng J.~L.,  Kaplinghat M.,  Tu H.,   Yu H.-B.,  2009, \mn@doi [JCAP]
  {10.1088/1475-7516/2009/07/004}, 0907, 004

\bibitem[\protect\citeauthoryear{Feng, Kaplinghat  \& Yu}{Feng
  et~al.}{2010}]{Feng:2009hw}
Feng J.~L.,  Kaplinghat M.,   Yu H.-B.,  2010, \mn@doi [Phys. Rev. Lett.]
  {10.1103/PhysRevLett.104.151301}, 104, 151301

\bibitem[\protect\citeauthoryear{Firmani, D'Onghia, Chincarini, Hernandez  \&
  Avila-Reese}{Firmani et~al.}{2001}]{Firmani:2000qe}
Firmani C.,  D'Onghia E.,  Chincarini G.,  Hernandez X.,   Avila-Reese V.,
  2001, \mn@doi [Mon. Not. Roy. Astron. Soc.]
  {10.1046/j.1365-8711.2001.04030.x}, 321, 713

\bibitem[\protect\citeauthoryear{{Gehrels}}{{Gehrels}}{1986}]{Gehrels:1986}
{Gehrels} N.,  1986, \mn@doi [\apj] {10.1086/164079}, \href
  {http://adsabs.harvard.edu/abs/1986ApJ...303..336G} {303, 336}

\bibitem[\protect\citeauthoryear{Gnedin \& Ostriker}{Gnedin \&
  Ostriker}{2001}]{Gnedin:2000ea}
Gnedin O.~Y.,  Ostriker J.~P.,  2001, \mn@doi [Astrophys. J.] {10.1086/323211},
  561, 61

\bibitem[\protect\citeauthoryear{Hahn \& Abel}{Hahn \&
  Abel}{2011}]{Hahn:2011uy}
Hahn O.,  Abel T.,  2011, \mn@doi [Mon. Not. Roy. Astron. Soc.]
  {10.1111/j.1365-2966.2011.18820.x}, 415, 2101

\bibitem[\protect\citeauthoryear{Hansen \& Moore}{Hansen \&
  Moore}{2006}]{Hansen:2004qs}
Hansen S.~H.,  Moore B.,  2006, \mn@doi [New Astron.]
  {10.1016/j.newast.2005.09.001}, 11, 333

\bibitem[\protect\citeauthoryear{Hansen \& Piffaretti}{Hansen \&
  Piffaretti}{2007}]{Hansen:2007jj}
Hansen S.~H.,  Piffaretti R.,  2007, \mn@doi [Astron. Astrophys.]
  {10.1051/0004-6361:20078656}, 476, L37

\bibitem[\protect\citeauthoryear{Hansen, Juncher  \& Sparre}{Hansen
  et~al.}{2010}]{Hansen:2010kb}
Hansen S.~H.,  Juncher D.,   Sparre M.,  2010, \mn@doi [Astrophys. J.]
  {10.1088/2041-8205/718/2/L68}, 718, L68

\bibitem[\protect\citeauthoryear{{Harvey}, {Massey}, {Kitching}, {Taylor}  \&
  {Tittley}}{{Harvey} et~al.}{2015}]{Harvey:2015}
{Harvey} D.,  {Massey} R.,  {Kitching} T.,  {Taylor} A.,   {Tittley} E.,  2015,
  \mn@doi [Science] {10.1126/science.1261381}, \href
  {http://adsabs.harvard.edu/abs/2015Sci...347.1462H} {347, 1462}

\bibitem[\protect\citeauthoryear{Harvey, Robertson, Massey  \& Kneib}{Harvey
  et~al.}{2016}]{Harvey:2016bqd}
Harvey D.,  Robertson A.,  Massey R.,   Kneib J.-P.,  2016, ]
  {10.1093/mnras/stw2671}

\bibitem[\protect\citeauthoryear{Host, Hansen, Piffaretti, Morandi, Ettori, Kay
   \& Valdarnini}{Host et~al.}{2009}]{Host:2008gi}
Host O.,  Hansen S.~H.,  Piffaretti R.,  Morandi A.,  Ettori S.,  Kay S.~T.,
  Valdarnini R.,  2009, \mn@doi [Astrophys. J.] {10.1088/0004-637X/690/1/358},
  690, 358

\bibitem[\protect\citeauthoryear{Kamada, Kaplinghat, Pace  \& Yu}{Kamada
  et~al.}{2016}]{Kamada:2016euw}
Kamada A.,  Kaplinghat M.,  Pace A.~B.,   Yu H.-B.,  2016

\bibitem[\protect\citeauthoryear{{Kaplinghat}, {Tulin}  \& {Yu}}{{Kaplinghat}
  et~al.}{2016}]{Kaplinghat:2016}
{Kaplinghat} M.,  {Tulin} S.,   {Yu} H.-B.,  2016, \mn@doi [Physical Review
  Letters] {10.1103/PhysRevLett.116.041302}, \href
  {http://adsabs.harvard.edu/abs/2016PhRvL.116d1302K} {116, 041302}

\bibitem[\protect\citeauthoryear{Kim, Peter  \& Wittman}{Kim
  et~al.}{2017}]{Kim:2016ujt}
Kim S.~Y.,  Peter A. H.~G.,   Wittman D.,  2017, \mn@doi [Mon. Not. Roy.
  Astron. Soc.] {10.1093/mnras/stx896}, 469, 1414

\bibitem[\protect\citeauthoryear{Lemze et~al.,}{Lemze
  et~al.}{2012}]{Lemze:2011ud}
Lemze D.,  et~al., 2012, \mn@doi [Astrophys. J.] {10.1088/0004-637X/752/2/141},
  752, 141

\bibitem[\protect\citeauthoryear{Loeb \& Weiner}{Loeb \&
  Weiner}{2011}]{Loeb:2010gj}
Loeb A.,  Weiner N.,  2011, \mn@doi [Phys. Rev. Lett.]
  {10.1103/PhysRevLett.106.171302}, 106, 171302

\bibitem[\protect\citeauthoryear{Ludlow et~al.,}{Ludlow
  et~al.}{2013}]{Ludlow:2013bd}
Ludlow A.~D.,  et~al., 2013, \mn@doi [Mon. Not. Roy. Astron. Soc.]
  {10.1093/mnras/stt526}, 432, 1103

\bibitem[\protect\citeauthoryear{Navarro, Frenk  \& White}{Navarro
  et~al.}{1997}]{Navarro:1996gj}
Navarro J.~F.,  Frenk C.~S.,   White S. D.~M.,  1997, \mn@doi [Astrophys. J.]
  {10.1086/304888}, 490, 493

\bibitem[\protect\citeauthoryear{Newman, Treu, Ellis  \& Sand}{Newman
  et~al.}{2013}]{Newman:2012nw}
Newman A.~B.,  Treu T.,  Ellis R.~S.,   Sand D.~J.,  2013, \mn@doi [Astrophys.
  J.] {10.1088/0004-637X/765/1/25}, 765, 25

\bibitem[\protect\citeauthoryear{Onorbe, Garrison-Kimmel, Maller, Bullock,
  Rocha  \& Hahn}{Onorbe et~al.}{2014}]{Onorbe:2013fpa}
Onorbe J.,  Garrison-Kimmel S.,  Maller A.~H.,  Bullock J.~S.,  Rocha M.,
  Hahn O.,  2014, \mn@doi [Mon. Not. Roy. Astron. Soc.]
  {10.1093/mnras/stt2020}, 437, 1894

\bibitem[\protect\citeauthoryear{{Peirani} et~al.,}{{Peirani}
  et~al.}{2016}]{Peirani:2016}
{Peirani} S.,  et~al., 2016, preprint, \href
  {http://adsabs.harvard.edu/abs/2016arXiv161109922P} {} (\mn@eprint {arXiv}
  {1611.09922})

\bibitem[\protect\citeauthoryear{Peter, Rocha, Bullock  \& Kaplinghat}{Peter
  et~al.}{2013}]{Peter:2012jh}
Peter A. H.~G.,  Rocha M.,  Bullock J.~S.,   Kaplinghat M.,  2013, \mn@doi
  [Mon. Not. Roy. Astron. Soc.] {10.1093/mnras/sts535}, 430, 105

\bibitem[\protect\citeauthoryear{Power, Navarro, Jenkins, Frenk, White,
  Springel, Stadel  \& Quinn}{Power et~al.}{2003}]{Power:2002sw}
Power C.,  Navarro J.~F.,  Jenkins A.,  Frenk C.~S.,  White S. D.~M.,  Springel
  V.,  Stadel J.,   Quinn T.~R.,  2003, \mn@doi [Mon. Not. Roy. Astron. Soc.]
  {10.1046/j.1365-8711.2003.05925.x}, 338, 14

\bibitem[\protect\citeauthoryear{Robertson, Massey, Eke  \& Bower}{Robertson
  et~al.}{2015}]{Robertson:2015faa}
Robertson A.,  Massey R.,  Eke V.,   Bower R.,  2015, \mn@doi [Mon. Not. Roy.
  Astron. Soc.] {10.1093/mnras/stv1805}, 453, 2267

\bibitem[\protect\citeauthoryear{{Robertson}, {Massey}  \& {Eke}}{{Robertson}
  et~al.}{2017}]{Robertson:2017}
{Robertson} A.,  {Massey} R.,   {Eke} V.,  2017, \mn@doi [\mnras]
  {10.1093/mnras/stw2670}, \href
  {http://adsabs.harvard.edu/abs/2017MNRAS.465..569R} {465, 569}

\bibitem[\protect\citeauthoryear{{Robles} et~al.,}{{Robles}
  et~al.}{2017}]{Robles2017}
{Robles} V.~H.,  et~al., 2017, preprint, \href
  {http://adsabs.harvard.edu/abs/2017arXiv170607514R} {} (\mn@eprint {arXiv}
  {1706.07514})

\bibitem[\protect\citeauthoryear{Rocha, Peter, Bullock, Kaplinghat,
  Garrison-Kimmel, Onorbe  \& Moustakas}{Rocha et~al.}{2013}]{Rocha:2012jg}
Rocha M.,  Peter A. H.~G.,  Bullock J.~S.,  Kaplinghat M.,  Garrison-Kimmel S.,
   Onorbe J.,   Moustakas L.~A.,  2013, \mn@doi [Mon. Not. Roy. Astron. Soc.]
  {10.1093/mnras/sts514}, 430, 81

\bibitem[\protect\citeauthoryear{{Schaller} et~al.,}{{Schaller}
  et~al.}{2015}]{Schaller:2015}
{Schaller} M.,  et~al., 2015, \mn@doi [\mnras] {10.1093/mnras/stv1067}, \href
  {http://adsabs.harvard.edu/abs/2015MNRAS.451.1247S} {451, 1247}

\bibitem[\protect\citeauthoryear{Schaye et~al.}{Schaye
  et~al.}{2015}]{Schaye:2014tpa}
Schaye J.,  et~al., 2015, \mn@doi [Mon. Not. Roy. Astron. Soc.]
  {10.1093/mnras/stu2058}, 446, 521

\bibitem[\protect\citeauthoryear{{Sparre} \& {Hansen}}{{Sparre} \&
  {Hansen}}{2012}]{Sparre:2012}
{Sparre} M.,  {Hansen} S.~H.,  2012, \mn@doi [\jcap]
  {10.1088/1475-7516/2012/10/049}, \href
  {http://adsabs.harvard.edu/abs/2012JCAP...10..049S} {10, 049}

\bibitem[\protect\citeauthoryear{Spergel \& Steinhardt}{Spergel \&
  Steinhardt}{2000}]{Spergel:1999mh}
Spergel D.~N.,  Steinhardt P.~J.,  2000, \mn@doi [Phys. Rev. Lett.]
  {10.1103/PhysRevLett.84.3760}, 84, 3760

\bibitem[\protect\citeauthoryear{Springel}{Springel}{2010}]{Springel:2009aa}
Springel V.,  2010, \mn@doi [Mon. Not. Roy. Astron. Soc.]
  {10.1111/j.1365-2966.2009.15715.x}, 401, 791

\bibitem[\protect\citeauthoryear{Springel, White, Tormen  \&
  Kauffmann}{Springel et~al.}{2001}]{Springel:2000qu}
Springel V.,  White S. D.~M.,  Tormen G.,   Kauffmann G.,  2001, \mn@doi [Mon.
  Not. Roy. Astron. Soc.] {10.1046/j.1365-8711.2001.04912.x}, 328, 726

\bibitem[\protect\citeauthoryear{{Springel} et~al.,}{{Springel}
  et~al.}{2005}]{Springel:2005}
{Springel} V.,  et~al., 2005, \mn@doi [\nat] {10.1038/nature03597}, \href
  {http://adsabs.harvard.edu/abs/2005Natur.435..629S} {435, 629}

\bibitem[\protect\citeauthoryear{Springel et~al.,}{Springel
  et~al.}{2008}]{Springel:2008cc}
Springel V.,  et~al., 2008, \mn@doi [Mon. Not. Roy. Astron. Soc.]
  {10.1111/j.1365-2966.2008.14066.x}, 391, 1685

\bibitem[\protect\citeauthoryear{{Suto}, {Peirani}, {Dubois}, {Kitayama},
  {Nishimichi}, {Sasaki}  \& {Suto}}{{Suto} et~al.}{2017}]{Suto:2017}
{Suto} D.,  {Peirani} S.,  {Dubois} Y.,  {Kitayama} T.,  {Nishimichi} T.,
  {Sasaki} S.,   {Suto} Y.,  2017, \mn@doi [\pasj] {10.1093/pasj/psw118}, \href
  {http://adsabs.harvard.edu/abs/2017PASJ...69...14S} {69, 14}

\bibitem[\protect\citeauthoryear{Tulin, Yu  \& Zurek}{Tulin
  et~al.}{2012}]{Tulin:2012re}
Tulin S.,  Yu H.-B.,   Zurek K.~M.,  2012, \mn@doi [JCAP]
  {10.1088/1475-7516/2012/05/013}, 1205, 013

\bibitem[\protect\citeauthoryear{Vogelsberger \& Zavala}{Vogelsberger \&
  Zavala}{2013}]{Vogelsberger:2012sa}
Vogelsberger M.,  Zavala J.,  2013, \mn@doi [Mon. Not. Roy. Astron. Soc.]
  {10.1093/mnras/sts712}, 430, 1722

\bibitem[\protect\citeauthoryear{Vogelsberger, Zavala  \& Loeb}{Vogelsberger
  et~al.}{2012}]{Vogelsberger:2012ku}
Vogelsberger M.,  Zavala J.,   Loeb A.,  2012, \mn@doi [Mon. Not. Roy. Astron.
  Soc.] {10.1111/j.1365-2966.2012.21182.x}, 423, 3740

\bibitem[\protect\citeauthoryear{{Vogelsberger} et~al.,}{{Vogelsberger}
  et~al.}{2014a}]{2014MNRAS.444.1518V}
{Vogelsberger} M.,  et~al., 2014a, \mn@doi [\mnras] {10.1093/mnras/stu1536},
  \href {http://adsabs.harvard.edu/abs/2014MNRAS.444.1518V} {444, 1518}

\bibitem[\protect\citeauthoryear{Vogelsberger, Zavala, Simpson  \&
  Jenkins}{Vogelsberger et~al.}{2014b}]{Vogelsberger:2014pda}
Vogelsberger M.,  Zavala J.,  Simpson C.,   Jenkins A.,  2014b, \mn@doi [Mon.
  Not. Roy. Astron. Soc.] {10.1093/mnras/stu1713}, 444, 3684

\bibitem[\protect\citeauthoryear{{Vogelsberger} et~al.,}{{Vogelsberger}
  et~al.}{2014c}]{2014Natur.509..177V}
{Vogelsberger} M.,  et~al., 2014c, \mn@doi [\nat] {10.1038/nature13316}, \href
  {http://adsabs.harvard.edu/abs/2014Natur.509..177V} {509, 177}

\bibitem[\protect\citeauthoryear{Vogelsberger, Zavala, Cyr-Racine, Pfrommer,
  Bringmann  \& Sigurdson}{Vogelsberger et~al.}{2016}]{Vogelsberger:2015gpr}
Vogelsberger M.,  Zavala J.,  Cyr-Racine F.-Y.,  Pfrommer C.,  Bringmann T.,
  Sigurdson K.,  2016, \mn@doi [Mon. Not. Roy. Astron. Soc.]
  {10.1093/mnras/stw1076}, 460, 1399

\bibitem[\protect\citeauthoryear{{Wittman}, {Golovich}  \& {Dawson}}{{Wittman}
  et~al.}{2017}]{Wittman:2017}
{Wittman} D.,  {Golovich} N.,   {Dawson} W.~A.,  2017, preprint, \href
  {http://adsabs.harvard.edu/abs/2017arXiv170105877W} {} (\mn@eprint {arXiv}
  {1701.05877})

\bibitem[\protect\citeauthoryear{Wojtak, Gottloeber  \& Klypin}{Wojtak
  et~al.}{2013}]{Wojtak:2013eia}
Wojtak R.,  Gottloeber S.,   Klypin A.,  2013, \mn@doi [Mon. Not. Roy. Astron.
  Soc.] {10.1093/mnras/stt1113}, 434, 1576

\bibitem[\protect\citeauthoryear{Yoshida, Springel, White  \& Tormen}{Yoshida
  et~al.}{2000}]{Yoshida:2000uw}
Yoshida N.,  Springel V.,  White S. D.~M.,   Tormen G.,  2000, \mn@doi
  [Astrophys. J.] {10.1086/317306}, 544, L87

\bibitem[\protect\citeauthoryear{Zavala, Vogelsberger  \& Walker}{Zavala
  et~al.}{2013}]{Zavala:2012us}
Zavala J.,  Vogelsberger M.,   Walker M.~G.,  2013, \mn@doi [Monthly Notices of
  the Royal Astronomical Society: Letters] {10.1093/mnrasl/sls053}, 431, L20

\bibitem[\protect\citeauthoryear{Zemp, Moore, Stadel, Carollo  \& Madau}{Zemp
  et~al.}{2008}]{Zemp:2007nt}
Zemp M.,  Moore B.,  Stadel J.,  Carollo C.~M.,   Madau P.,  2008, \mn@doi
  [Mon. Not. Roy. Astron. Soc.] {10.1111/j.1365-2966.2008.13126.x}, 386, 1543

\bibitem[\protect\citeauthoryear{Zemp, Gnedin, Gnedin  \& Kravtsov}{Zemp
  et~al.}{2011}]{Zemp:2011ed}
Zemp M.,  Gnedin O.~Y.,  Gnedin N.~Y.,   Kravtsov A.~V.,  2011, \mn@doi
  [Astrophys. J. Suppl.] {10.1088/0067-0049/197/2/30}, 197, 30

\bibitem[\protect\citeauthoryear{van~den Aarssen, Bringmann  \&
  Pfrommer}{van~den Aarssen et~al.}{2012}]{Aarssen:2012fx}
van~den Aarssen L.~G.,  Bringmann T.,   Pfrommer C.,  2012, \mn@doi [Phys. Rev.
  Lett.] {10.1103/PhysRevLett.109.231301}, 109, 231301

\makeatother
\end{thebibliography}

\appendix
\section{Halo sample and convergence}
\begin{figure}
	\centering
	\includegraphics[width=0.49\linewidth]{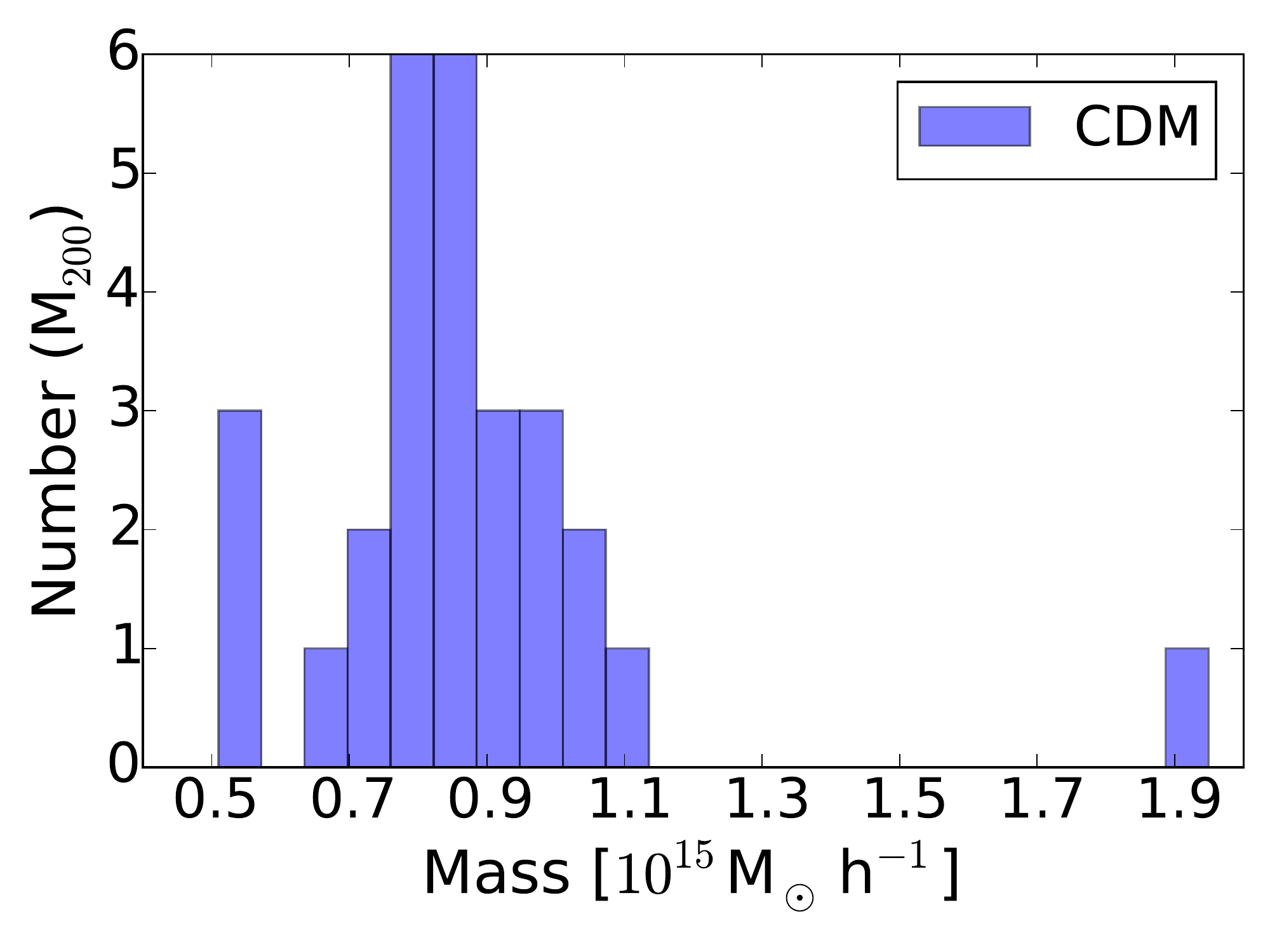}
	\includegraphics[width=0.49\linewidth]{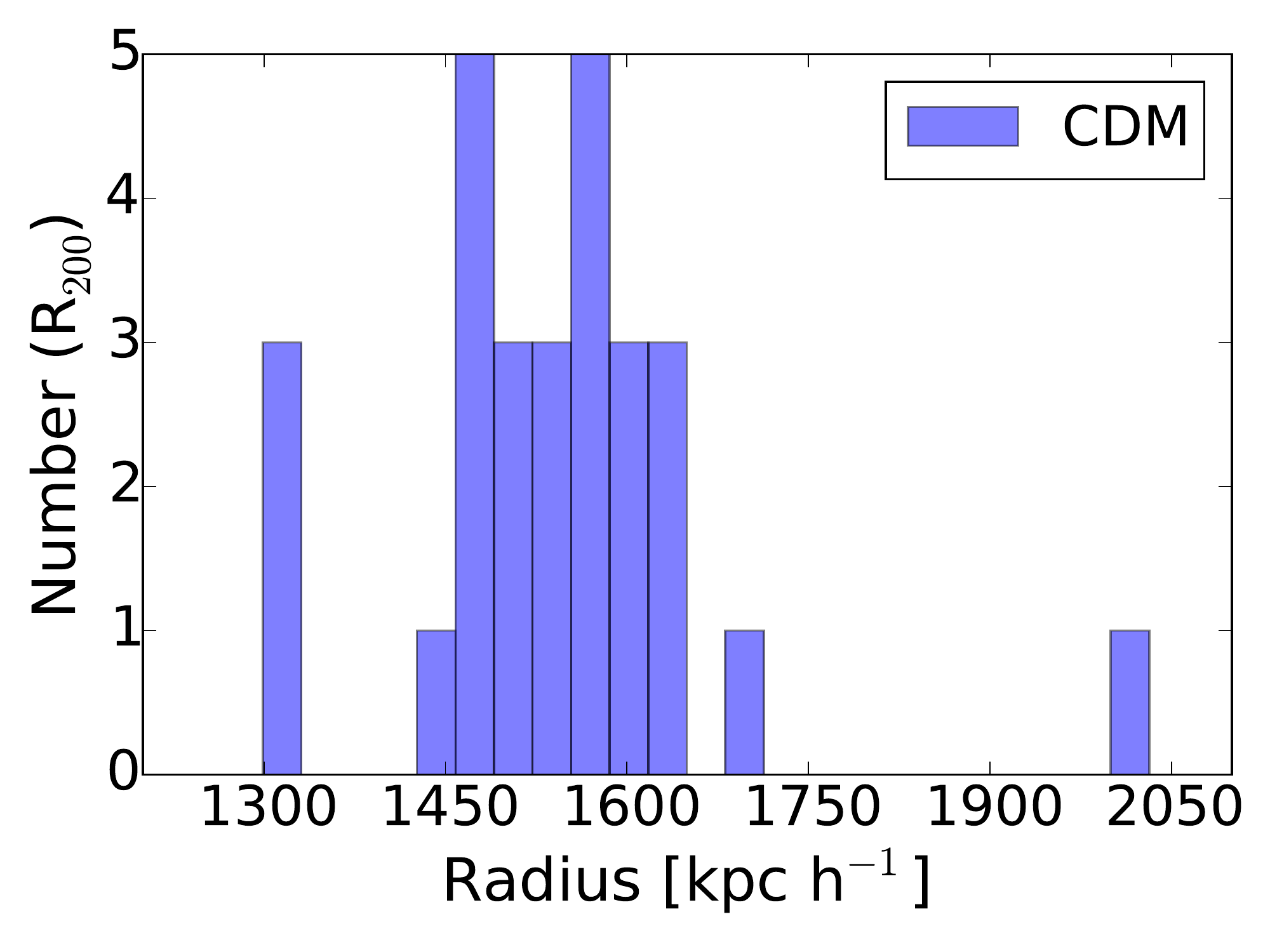}
	\caption{The distribution of virial masses and radii for the 28 zoom resimulated CDM halos used in this work. \textbf{Left}: Number of halos vs virial mass. Except for one outlier near 1.9 $\times 10^{15}$ M$_{\astrosun}$ h$^{-1}$, we have a heavily peaked distribution around 0.9 $\times 10^{15}$ M$_{\astrosun}$ h$^{-1}$. \textbf{Right}: Number of halos vs virial radius. We see a peaked distribution around 1550 kpc h$^{-1}$, with the same halo as before showing up as an outlier at near 2000 kpc h$^{-1}$. For the SIDM 1 and SIDM 0.1 halos (not shown) the distributions are virtually the same, because SIDM does not significantly alter the virial mass or radius of a halo.}
	\label{fig:sims:info}
\end{figure}

Figure \ref{fig:sims:info} shows the distributions in M$_{200}$ and R$_{200}$ for the 28 halos zoom resimulations for this paper. Except for the most massive cluster, the sample has a narrow distribution centered around M$_{200} \sim 0.9 \times 10^{15}$ M$_{\astrosun}$ h$^{-1}$ and R$_{200} \sim 1550$ kpc h$^{-1}$.

\subsection{Halo shapes and density profiles}
\label{sec:convergence}
We followed the discussion in \citet{Onorbe:2013fpa} for our tests and took their recommendation to define the Lagrangian volume as the ellipse enclosing all the particles at $z=50$ that were inside the spherical radius $R_{tb} = (1.5 \Delta_{res} + 7) \times R_{200}$ at $z=0$, where $\Delta_{res}$ is the resolution difference between the parent and zoom simulations, which in our case is 4. We tested the impact of varying the Lagrangian volume between $\Delta_{res}=2.5-5$ (and shape, which made little difference) for the most massive halo in our sample while keeping the resolution fixed (corresponding to 4096$^3$ particles for the high resolution region) and found that for our purposes the smaller volume of $\Delta_{res}=3$ provided sufficient convergence for the density profiles, at the percent level at the Power radius (defined in section \ref{sec:shapes}).
	
Similarly, we estimated the convergence of our simulations by varying the resolution of the zoom region of the simulation, using a number of particles corresponding to $2^{\rm n}$ cubed for the high resolution region, with the varying levels of resolution being $n=11$, 12 and 13. We found that the density profiles are converged to 4\% at a radius of four times the softening length and to a percent level at the Power radius.
	
For the halo shapes we performed the same tests as above and found that they are converged to $\sim$10\% at around 50~kpc, rapidly improving to $\sim$5\% at around 100~kpc and to around $1-3$\% level at larger radii.

We also checked the level of contamination across resolution levels, i.e., the number of lower resolution particles present in the high resolution region we end up analysing, which we found to be non existent to negligible.

\subsection{Velocity anisotropy}
\label{sec:conv:beta}
\begin{figure}
	\centering
	\includegraphics[width=\linewidth]{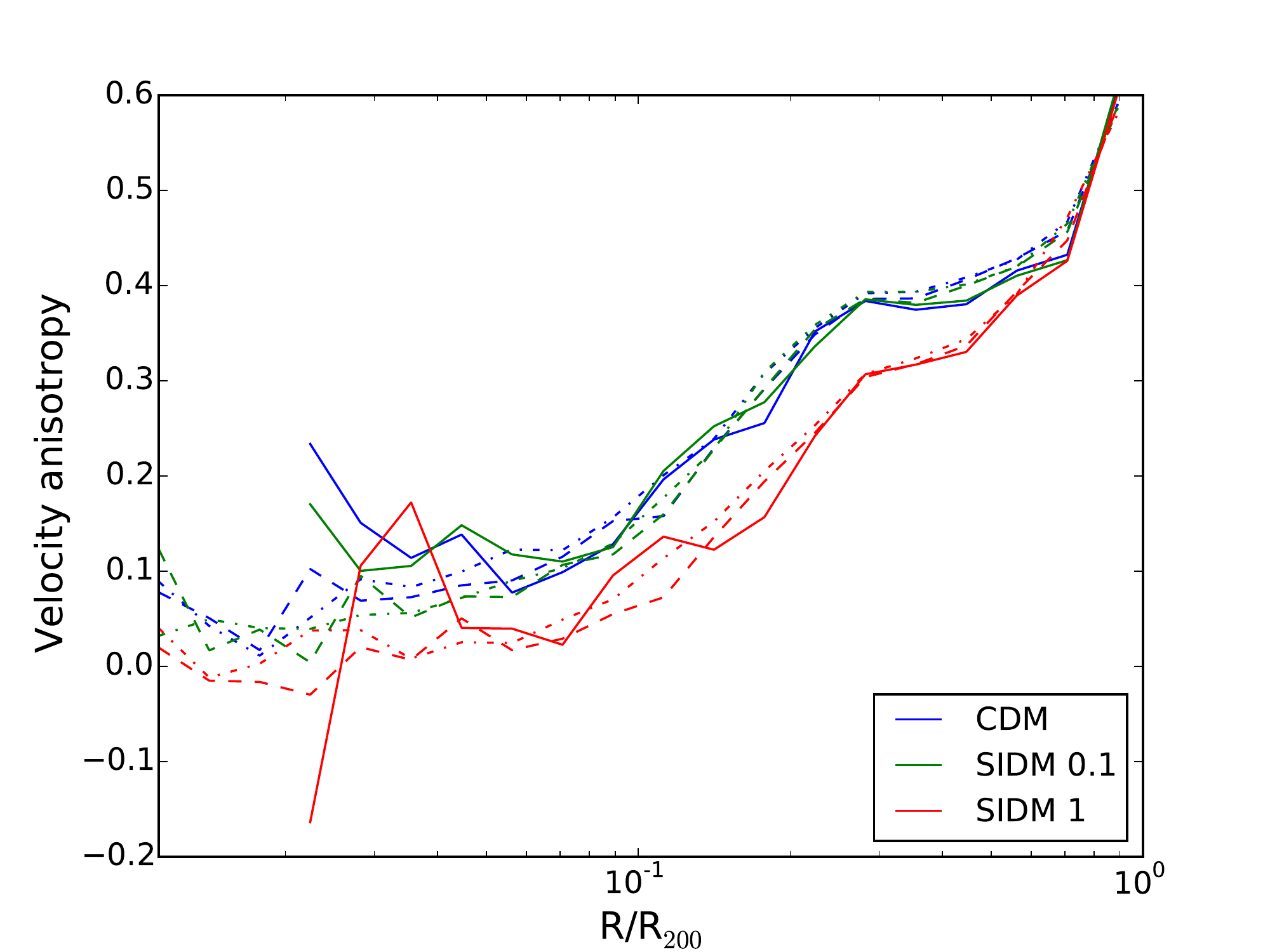}
	\caption{Convergence plots for the velocity anisotropy profile of our most massive halo, varying effective resolution between $2^{11}$ and $2^{13}$ particles. We have CDM in blue, SIDM0.1 in green and SIDM1 in red. Resolution level 11 is indicated by a solid line, level 12 by a dashed line and level 13 by a dotted line. For level 12, the resolution level used for our analysis, the profiles are fairly well converged down to around 0.04 R$_{200}$, which we take as our trust radius.}
	\label{fig:conv:beta}
\end{figure}

We compared beta profiles across multiple resolution levels ($n=11-13$, as above) for our most massive halo, in order to identify a trust radius, above which we trust the beta profile results. This is shown in Figure \ref{fig:conv:beta} and we find that we can trust the results above 0.04 $R_{200}$ for the resolution level ($n=12$) used for the analysis in this work.

\section{Halo shapes code testing}
\label{sec:code_testing}
The halo shapes code we used in this paper was tested using artificial haloes, generated with predefined density profiles and shapes. Each of these haloes has about 14 million particles, with M$_{\rm halo} = 10^{15}$M$_{\astrosun}$ and R$_{200} = 2000$ kpc. They were created using the same softening ($\epsilon = 5.425$ kpc/h) and cosmology ($\Omega_{\rm m} = 0.315$, $\Omega_{\Lambda} = 0.685$ and $h=0.673$) as for the main sample of the paper. The code reproduced the input shapes with sufficient precision as we describe below. The particle distributions within these test haloes were generated using Marcel Zemp's \texttt{HALOGEN} code\footnote{http://www.marcelzemp.ch/software.html} \citep{Zemp:2007nt}. The haloes cover a variety of shapes, close to spherical, prolate or oblate and with rapidly increasing or declining sphericity as a function of radius. We also explored density distributions close to those expected in CDM (cuspy) and SIDM (cored), by using density profiles given by
\begin{align}
\rho(r) = \frac{\rho_s}{(r/r_s)^{\gamma}(1+(r/r_s)^{\alpha})^{(\beta - \gamma)/\alpha}},
\label{eq:halogen}
\end{align}
where $\rho_s$ and $r_s$ are the scale density and radius, respectively. By setting $\alpha = 1$, $\beta = 3$ and $\gamma = 1$, Eq.~\ref{eq:halogen} reduces to the familiar NFW cuspy profile, whereas when setting $\gamma = 0$, we can explore cored density profiles.

We tested different options available in the halo shapes code, namely:
\begin{itemize}
	\item fix the magnitude of the major axis of a shell ($a$ fixed),
	\item fix the volume of the shell (V$_{\rm shell}$ fixed),
	\item shapes computed from the full enclosed ellipsoidal volume instead of the shell volume ($a, b, c$ for V$_{\rm encl}$ rather than V$_{\rm shell}$),
	\item weight the contribution of each particle to the shape tensor by the elliptical radius, $r_{\rm ell} = \sqrt{x_{\rm ell}^2 + y_{\rm ell}^2 \left(b/a\right)^{-2} + z_{\rm ell}^2 \left(c/a\right)^{-2}}$, where $x_{\rm ell}$, $y_{\rm ell}$ and $z_{\rm ell}$ are the distances along the principal axes.
\end{itemize}

\begin{figure}
	\centering
	\includegraphics[width=\linewidth]{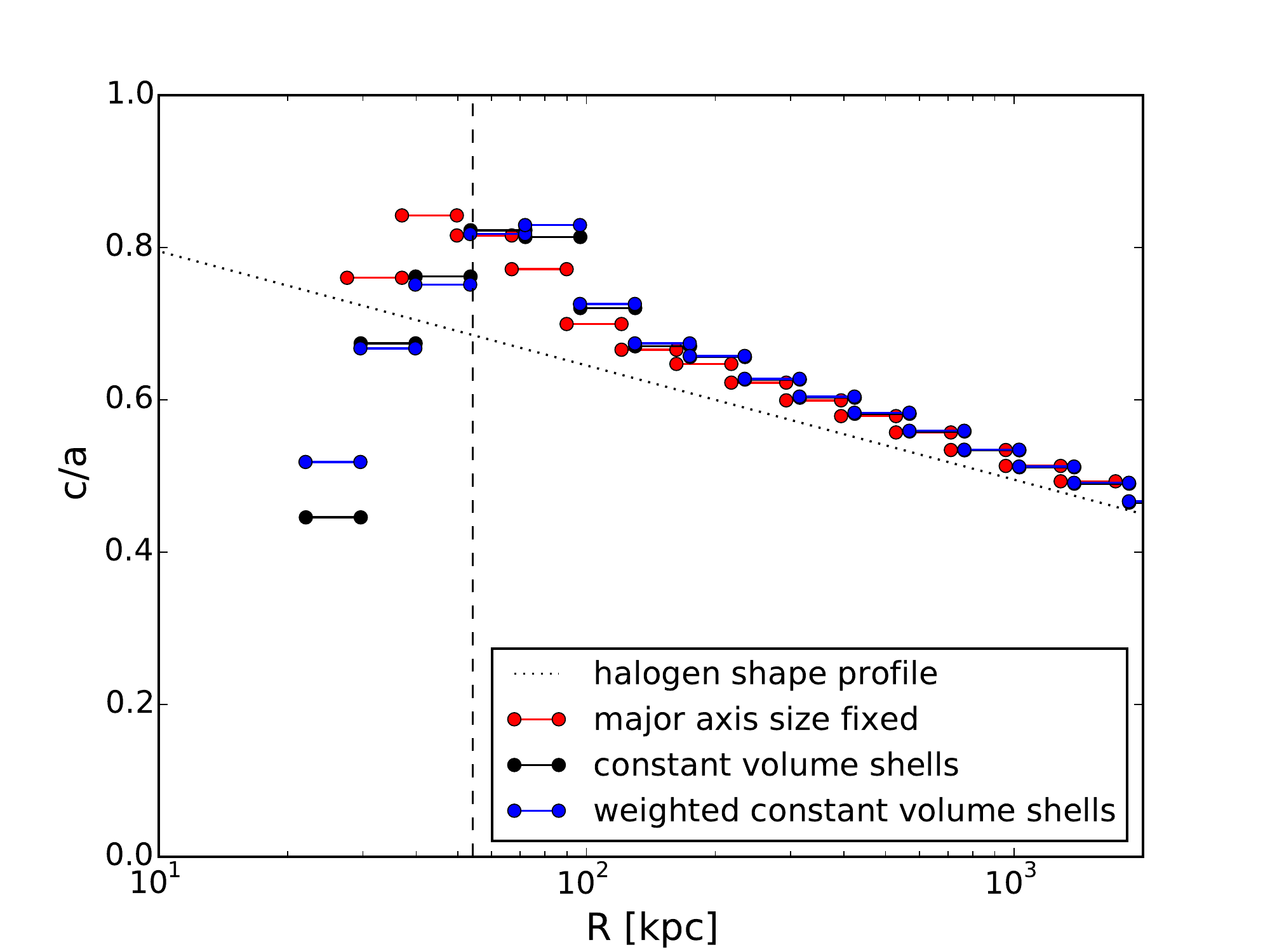}
	\caption{Comparison of the more accurate halo shape methods implemented in our code. We used a halo with a known shape profile, generated with \texttt{HALOGEN} \citep{Zemp:2007nt}. The figure shows the ratio of the minor (c) to major axis (a) vs radius, and the black dotted line indicates the analytical halo shape profile used to generate the halo with \texttt{HALOGEN} ($\gamma = 0$ was used for the density profile, corresponding to a cored profile, see eq. \ref{eq:halogen}). The vertical dashed line marks ten times the softening for this simulation, which is a measure for the radius above which the halo shapes are reliably converged. The overestimation of sphericity (especially at low radii) is discussed in Fig. \ref{fig:testhalos:gamma}. We see that the method where the size of the major axis is fixed (red bins) is slightly more accurate than the methods that keep the volume of each shell constant (black and blue bins). For this paper we use only the fixed major axis method.}
	\label{fig:testhalos:methods}
\end{figure}

An example of the performance of the code for the case of a cored profile ($\gamma=0$), using only the most accurate options, is shown in Figure \ref{fig:testhalos:methods}. The dotted line shows the input radial dependence for the axis ratio $c/a$, while the circles with different colors are the values found by the code for a given shell (with a size given by the segment connecting the corresponding pair of circles). The vertical dashed line indicates ten times the softening, which is the adopted measure for when the halo shapes code is reliably converged. In general we find that the method of keeping the size of the major axis fixed performs the best, and is the one we chose for our analysis.

\begin{figure}
	{\centering
		$\boldsymbol{\gamma = 1}$\par\medskip}
	\centering
	\includegraphics[width=0.49\linewidth]{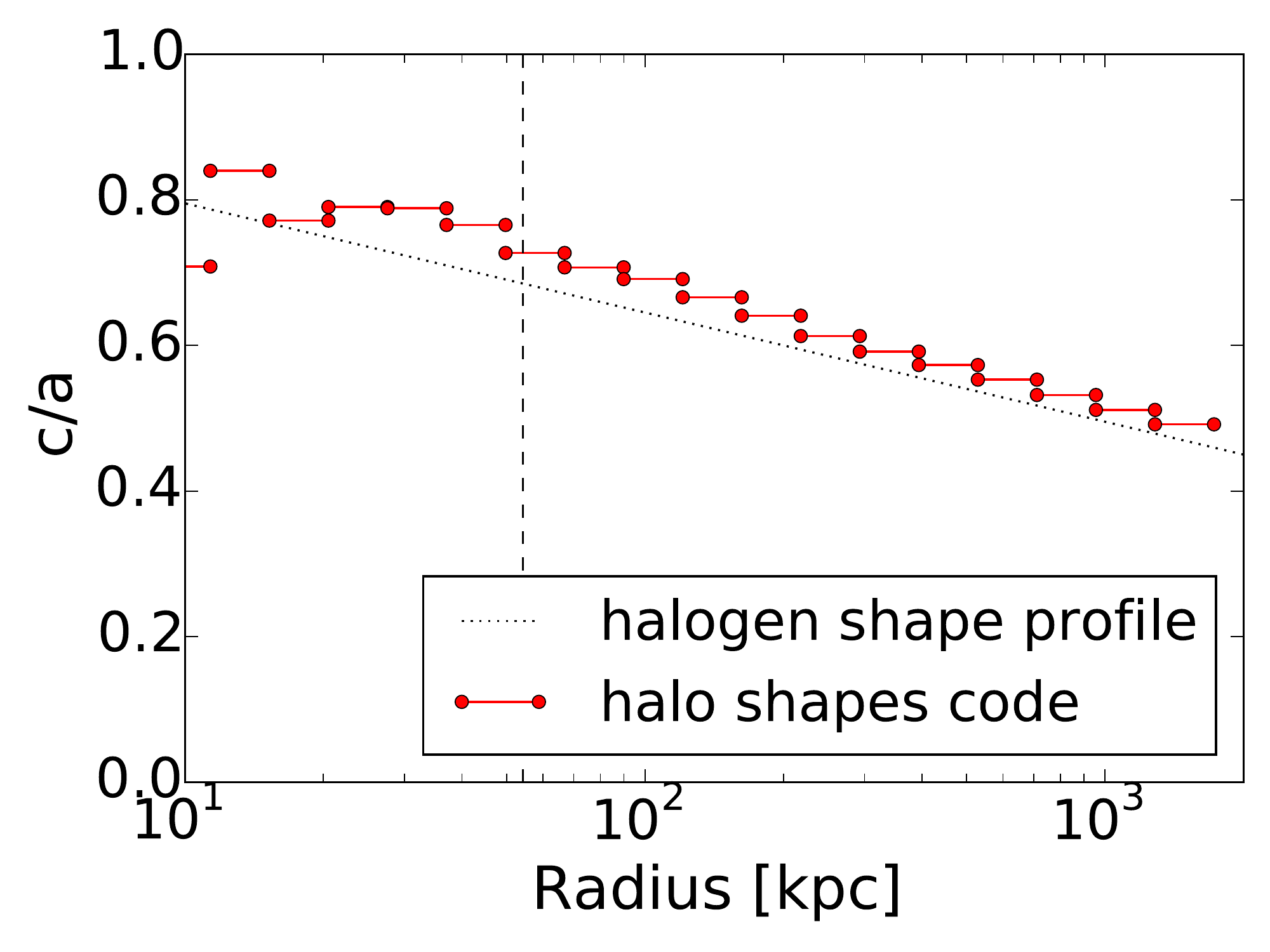}
	\includegraphics[width=0.49\linewidth]{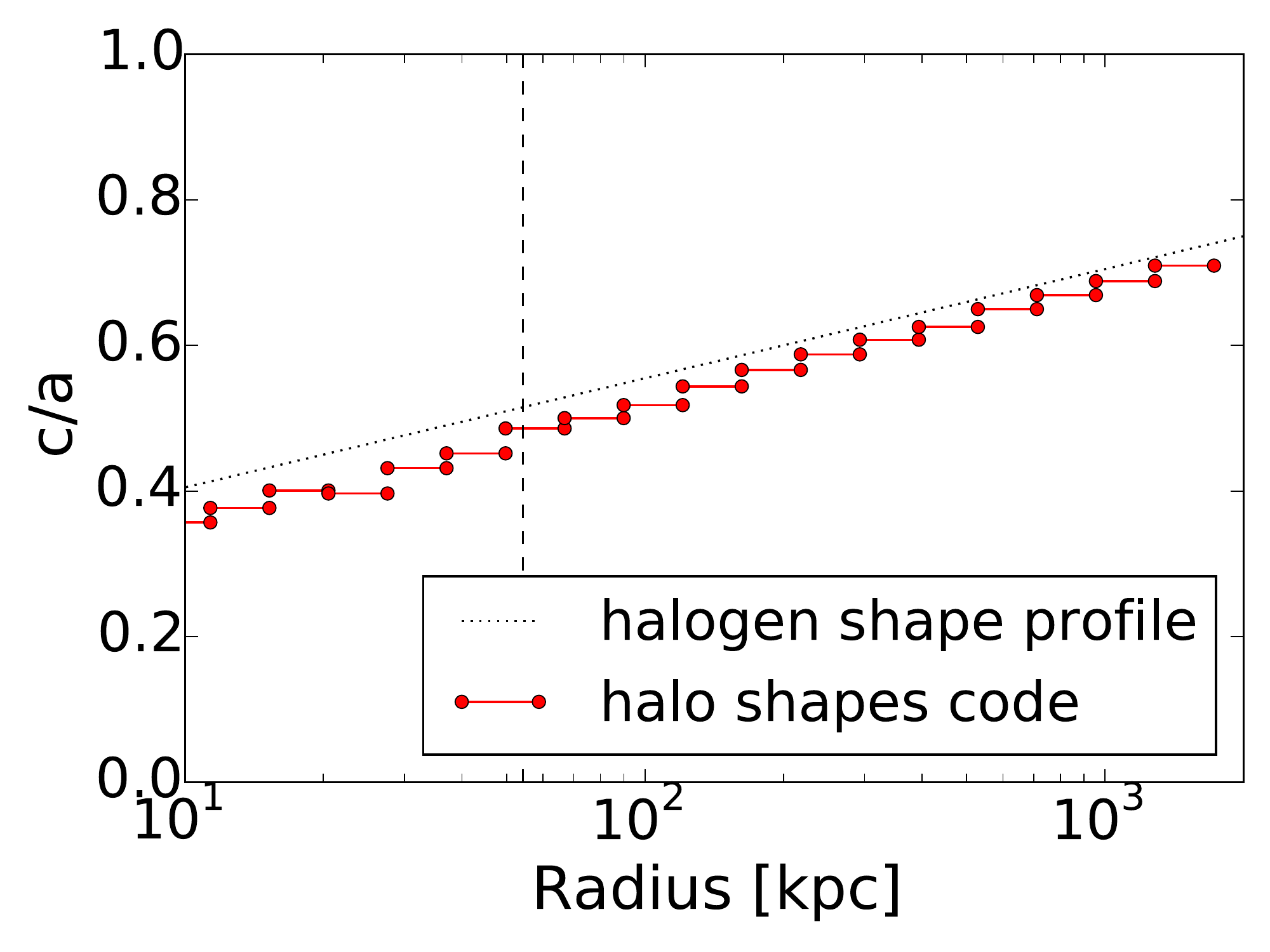}\\
	{\centering
		$\boldsymbol{\gamma = 0}$\par\medskip}
	\centering
	\includegraphics[width=0.49\linewidth]{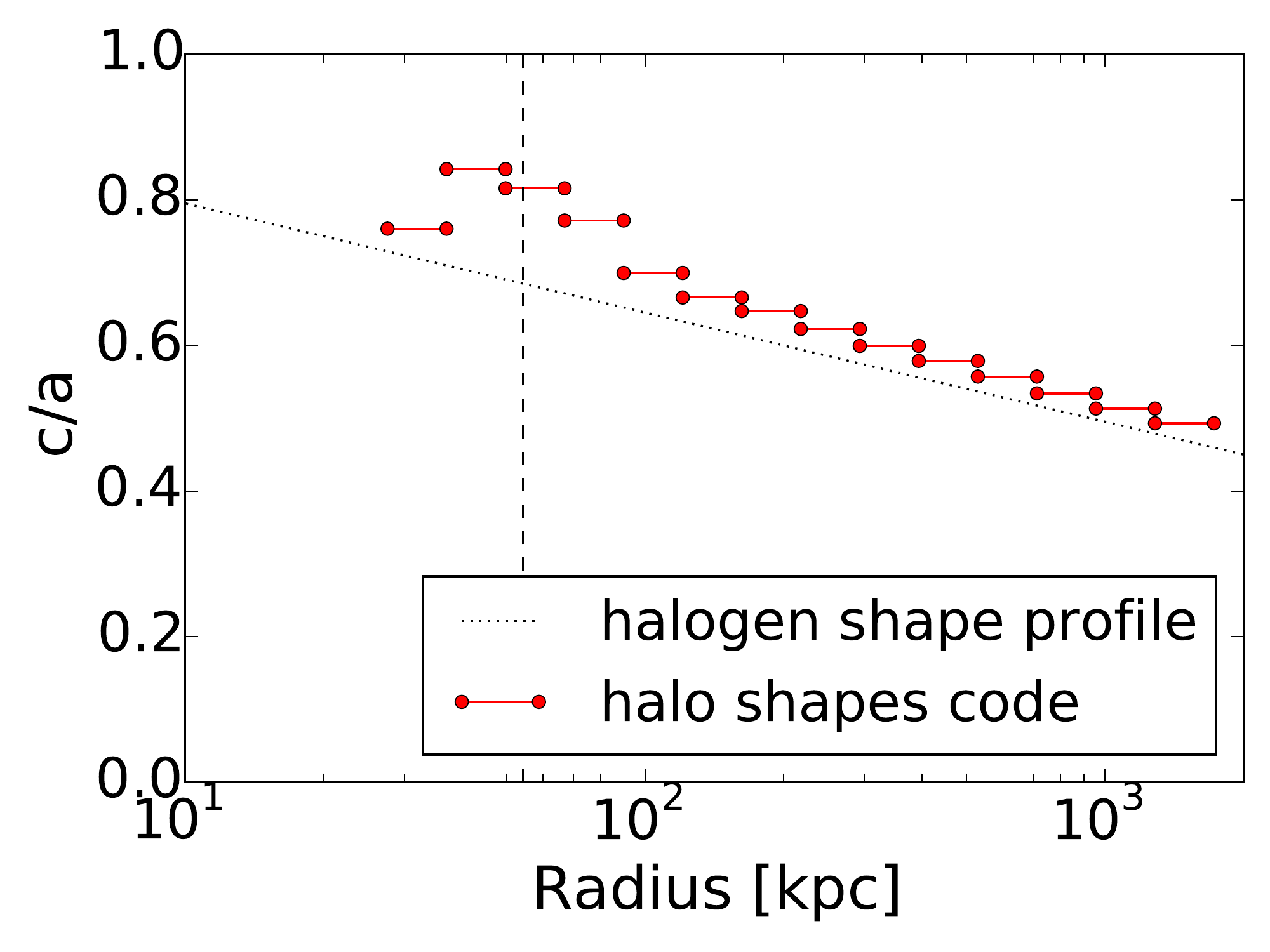}
	\includegraphics[width=0.49\linewidth]{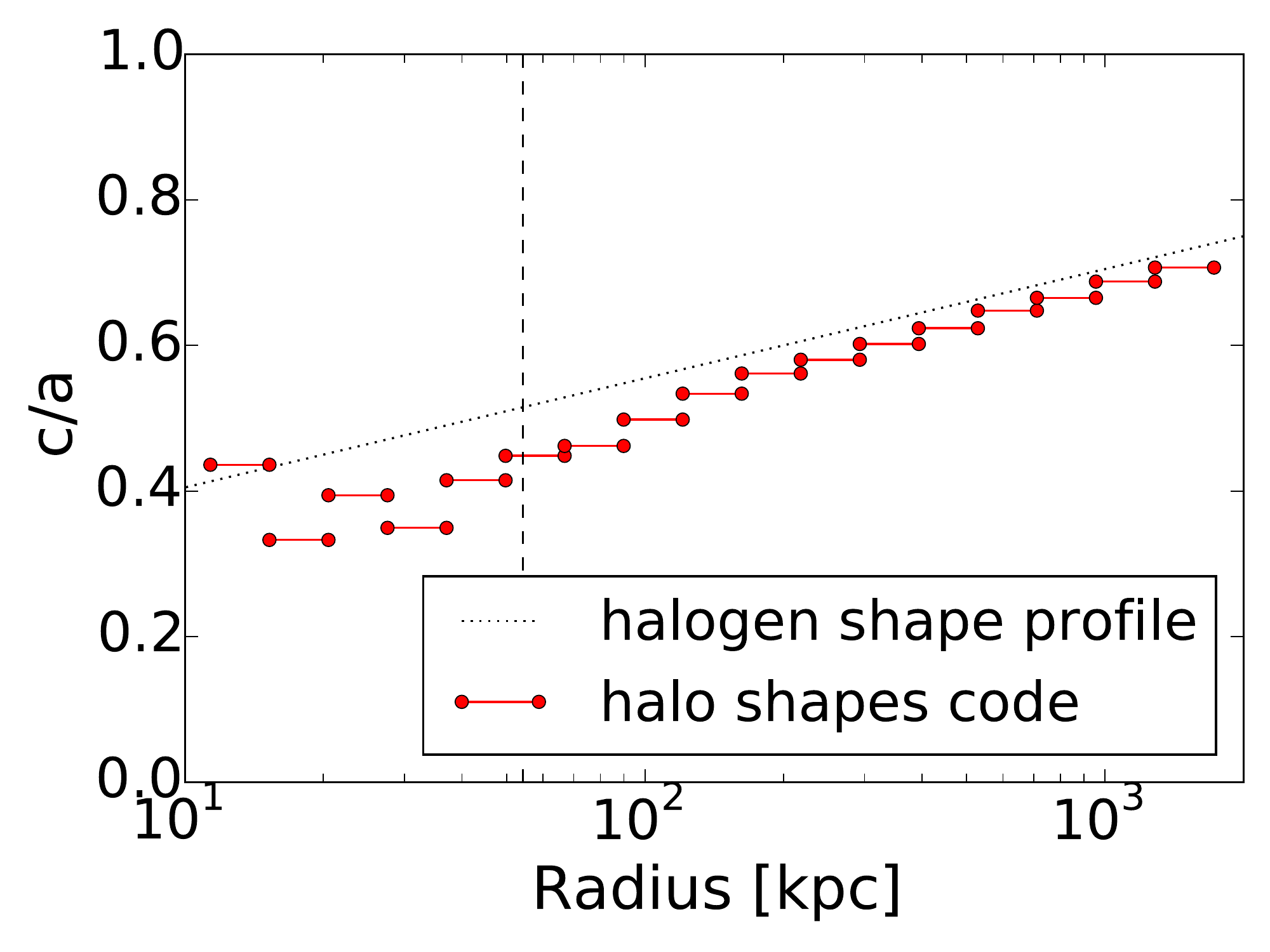}
	\caption{The halo shapes (ratio of minor to major axis) vs radius for four distinct haloes, two with cuspy central densities (top row with $\gamma=1$, see eq. \ref{eq:halogen}) and two with cored density profiles (bottom row with $\gamma=0$). The analytical halo shape profiles are shown by black dotted lines. The black dashed line indicates ten times the softening, which is a measure of the radius above which we can trust the halo shapes results, and the red bins show the results of our halo shapes code. For haloes declining (increasing) in sphericity with radius we see the halo shapes code slightly overestimates (underestimates) results, by $\sim 8\%$ in the top row and less than $\sim 10\%$ for the well converged bins in the bottom row. However, for the bottom row we see that due to the low number of particles at lower radii (because of the cored central densities of the haloes) the halo shapes code struggles to converge and overestimates the sphericity by as much as $\sim 20\%$.}
	\label{fig:testhalos:gamma}
\end{figure}

For shells where convergence is achieved, the performance of the code is good, but there is a clear bias towards higher or lower sphericity depending on the sign of the gradient of the input sphericity: the code overestimates, by about 8\% for well behaved bins, the axis ratios for haloes decreasing in sphericity with radius (see the left panel of Fig. \ref{fig:testhalos:gamma}) and slightly underestimates, by less than about 10\%, the axis ratios for haloes with increasing sphericity (see the right panel of Fig. \ref{fig:testhalos:gamma}). This bias gets progressively worse towards the center, due to a less steep local mass density profile and a reduction in the number of particles per shell, especially in combination with a high degree of sphericity. This means the performance of the code is less reliable towards the center of cored profiles, where the local mass density profile is nearly flat, compared to cuspy profiles with a steeper local density profile (compare the $\gamma=0$ profiles in the bottom panel of Fig. \ref{fig:testhalos:gamma} to the $\gamma = 1$ profiles in the top panel). As noted by \citet{Zemp:2011ed}, a difficulty in finding small density differences between shells is expected for a largely homogeneous distribution of particles. These authors also explained that the systematic bias is mostly due to the local mass density profile of the haloes. We confirm their results that increasing resolution and binning only slightly decreases this effect.

In realistic situations, the most severe limitation to the accuracy of the code is the particle number within a given ellipsoidal shape, as is clear by comparing the inner structure ($<100$ kpc) of the haloes in the top and bottom row of Fig. \ref{fig:testhalos:gamma}. The haloes in the top row ($\gamma=1$) have cuspy density profiles (applicable to CDM haloes) and thus more particles in the inner region, whereas the haloes in the bottom row ($\gamma=0$) have cored density profiles (applicable to SIDM haloes) and therefore fewer particles in the center. Obtaining convergence for shells with less than $\mathcal{O}(10^4)$ particles is often problematic and results are less accurate (or simply fail to converge). This is seen for the inner bins in the bottom row ($\gamma=0$) of Fig. \ref{fig:testhalos:gamma}, where the results are overestimated or underestimated by as much as about $20\%$. These biases must be kept in mind when interpreting the results of realistic SIDM haloes. For the analysis in this paper we formally require a minimum of 1500 particles for the innermost ellipsoidal shell, but due to the problems mentioned above for SIDM1 the low radius shells are not included, and, in any case, our analysis is focused on much larger radii.

\end{document}